\documentclass[3p]{elsarticle}
\usepackage{lineno,hyperref}
\usepackage{amsmath}
\usepackage{booktabs}
\usepackage{float}
\usepackage[normalem]{ulem}
\usepackage{color, soul}
\modulolinenumbers[1]
\usepackage{subfigure}
\usepackage{graphicx}
\usepackage{times}
\usepackage{epsfig}
\usepackage{graphicx}
\usepackage{amsmath}
\usepackage{amssymb}
\usepackage{graphics}
\usepackage{subfigure}
\usepackage{epstopdf}
\usepackage{array}
\usepackage{multirow}
\usepackage{color}
\usepackage{array}
\usepackage{booktabs}
\usepackage{comment}
\usepackage{dcolumn}
\usepackage{bm}
\usepackage{hyperref}
\usepackage{float}
\usepackage{multirow}
\usepackage{setspace}
\usepackage{url}
\usepackage{hyperref}
\usepackage{algorithm}  
\usepackage{algpseudocode}  
\usepackage{amsmath}

\usepackage{amssymb}

\begin{document}

\begin{frontmatter}



\title{A fast and flexible algorithm for microstructure reconstruction combining simulated annealing and deep learning}


\author[]{Zhenchuan Ma}

\author[]{Xiaohai He}

\author[]{Pengcheng Yan}

\author[]{Fan Zhang}

\author[]{Qizhi Teng\corref{cor}}
\ead{qzteng@scu.edu.cn}

\cortext[cor]{Corresponding author.}
\address{College of Electronics and Information Engineering, Sichuan University,  Chengdu 610065, China}

\begin{abstract}
The microstructure analyses of porous media have considerable research value for the study of macroscopic properties. As the premise of conducting these analyses, the accurate reconstruction of microstructure digital model is also an important component of the research. Computational reconstruction algorithms of microstructure have attracted much attention due to their low cost and excellent performance. However, it is still a challenge for computational reconstruction algorithms to achieve faster and more efficient reconstruction. The bottleneck lies in computational reconstruction algorithms, they are either too slow (traditional reconstruction algorithms) or not flexible to the training process (deep learning reconstruction algorithms). To address these limitations, we proposed a fast and flexible computational reconstruction algorithm, neural networks based on improved simulated annealing framework (ISAF-NN). The proposed algorithm is flexible and can complete training and reconstruction in a short time with only one two-dimensional image. By adjusting the size of input, it can also achieve reconstruction of arbitrary size. Finally, the proposed algorithm is experimentally performed on a variety of isotropic and anisotropic materials to verify the effectiveness and generalization.

\end{abstract}


\end{frontmatter}



\section{Introduction}

A porous medium is an aggregate composed of pore spaces and solid materials. As an essential  part of the microstructure of porous media, pore spaces have considerable research value for analyzing their performance, such as heat transfer performance \cite{1zhao2012review}, electrical properties \cite{2wang2013recent}, and seepage characteristics \cite{3lan2019review}.  In this regard, complex microscopic transport phenomena at the pore level are fundamental, as these phenomena influence macroscopic phenomena, such as increased heat transfer and pressure losses. Such analyses and studies have many applications in materials science, chemistry, water resources, petroleum geology, and other fields \cite{4chen2022pore}.

Accurate and rapid establishment of the digital model of porous structures  is the premise of conducting research and analyzing its performance, which can be performed by direct imaging or computational reconstruction methods. Direct imaging methods use imaging techniques, such as focused ion beam scanning electron microscopy (FIB-SEM) and X-ray computed tomography (XCT) \cite{5iwai2010quantification,6wildenschild2013x}, to directly image porous media. Its three-dimensional (3D) digital model is established by stacking the obtained scanning two-dimensional (2D) sequence images. However, obtaining 3D structures of porous media directly is difficult. Challenges to this method also include the imaging equipment, such as high equipment cost, complex operation, and a contradiction between the field of view and imaging resolution \cite{7tahmasebi2012reconstruction}. These factors limit the establishment of 3D digital models using direct imaging methods.

Compared with the direct imaging methods, computational reconstruction methods are faster, more efficient, and have a lower cost. Moreover, computational reconstruction methods can be used to obtain the porous media of the corresponding structure under specified parameters or constraint functions, which have a wide range of applications in microstructure characterization and reconstruction (MCR) \cite{8bostanabad2018computational,8_1sahimi2021reconstruction}. Therefore, MCR computational reconstruction methods have gained widespread attention in various fields of study. Currently, computational reconstruction methods primarily include classical optimization \cite{9yeong1998reconstructing}, pattern-matching \cite{10strebelle2002conditional,11mariethoz2010direct}, learning-based \cite{12mosser2017reconstruction}, and random field methods \cite{13yang2018new,14zhang2019efficient,15gao2021ultra,16guo2023spherical}. The first category is the classical optimization-based methods, such as  the simulated annealing (SA) method \cite{9yeong1998reconstructing}. Such methods use a statistical feature function and physical descriptors of the reference image as the optimization objective. First, a noisy 3D structure is initialized under certain conditions, such as the phase volume fraction of the reference image. Subsequently , the state of the 3D structure is updated by randomly swapping two points in different phases to attain optimization. Finally, optimization continues until the 3D structure attains the desired state and satisfies the objective.

Classical optimization-based methods extract a reference image's statistical features as the reconstruction's optimization objective. Effective features can be extracted from different reference images. Thus, such a method has strong generalization and applicability for reconstructing different porous media; however, the morphology of pore spaces in porous media cannot be effectively described using statistical functions. To accurately reproduce the morphology of the pore space, the pattern density function \cite{17gao2016pattern}, rescaled correlation functions \cite{18karsanina2018hierarchical}, co-occurrence correlation function \cite{19feng2018reconstruction}, fractal control function \cite{20zhou20183d}, and distance correlation functions \cite{21zhang2019high} have been proposed to characterize the pore space morphology and were used as the optimization objective. Corresponding hierarchical strategies \cite{22alexander2009hierarchical}, point selection strategies \cite{23tang2009pixel}, and other acceleration strategies \cite{24song2019improved,25xiao2022novel} have also been proposed to accelerate the iterative process. The same is true for descriptor-based reconstruction methods \cite{26yang2018new,27seibert2022descriptor,28bagherian2022new,29chen2022fast,30ajani2022microstructural}, and several improvements have been proposed to improve the speed and accuracy of the reconstruction.

Pattern-matching methods include multipoint statistics (MPS) methods. Classical MPS methods, such as single normal equation simulation (SNESIM) \cite{10strebelle2002conditional} and cross-correlation-based simulation (CCSIM) \cite{7tahmasebi2012reconstruction}, establish the pattern set by first extracting the multipoint statistics function of the reference image. Next, a specific point or patch to be reconstructed is found during the reconstruction, and the best matching pattern is located in the pattern set using the conditional data around the point or patch to be reconstructed. The corresponding position information in the best- matching pattern is assigned to the point or patch to be reconstructed according to specific criteria. Finally, the reconstruction process of the point or patch is repeated until the reconstruction structure is complete.

The multipoint statistic function provides a more accurate description of the morphology than the statistical feature function. MPS methods reproduce the reference image's multipoint statistic function, and the pore space's morphology is closer to the target structure than those in the SA methods. However, classical MPS methods are relatively inefficient owing to the number of pattern-matching processes and the parameter adjustments. To achieve a high CCSIM reconstruction efficiency, Tahmasebi et al. \cite{31tahmasebi2014ms} proposed a multi-scale strategy and accelerated the matching process using Fourier space computation. Pourfard et al. \cite{32pourfard2017pcto} proposed the use of parallel acceleration methods combined with optimization methods. Based on the direct sampling (DS) method \cite{11mariethoz2010direct}, Zuo et al. \cite{33zuo2020tree} proposed a clustering tree-based method for pattern search to improve reconstruction speed. In addition, other MPS methods \cite{34gao2015reconstruction,35gravey2020quicksampling,36bai2020hybrid,37wang2022two} have been proposed to improve the accuracy and speed of reconstruction.

Another major category is learning-based methods. Most of these methods are based on deep learning, whereas others are based on dictionary \cite{38li2018markov,39xia2021three} and machine learning \cite{40fu2022stochastic}. Methods based on deep learning mainly establish a mapping relationship between 2D images and 3D structures by training models with existing 3D structure information. The established mapping relationship is used to complete the reconstruction by inputting the reference image during the reconstruction. Mosser et al. \cite{12mosser2017reconstruction} introduced generative adversarial networks (GANs) for the 3D reconstruction of porous media; however, this method cannot reconstruct 3D structures from 2D reference images or target parameters, and its training process is unstable. Subsequent research addressed this issue and proposed various improved methods based on GANs \cite{41feng2018accelerating,42feng2020end,43valsecchi2020stochastic,44zhang20213d,45volkhonskiy2022generative,46zhang20223d,47henkes2022three}. In addition to GANs, other deep learning methods for 3D reconstruction have been proposed, such as transfer learning \cite{48bostanabad2020reconstruction}, deep neural networks \cite{49fu2021statistical}, recurrent neural networks \cite{50zhang20223d}, and flow models \cite{51anderson2020rockflow}.

The learning-based reconstruction method is fast, efficient, and widely studied. However, most deep learning methods require 3D structures as training samples, and model training requires time. This method cannot be generalized as effectively as traditional methods, such as SA and MPS, which can reconstruct the structure of various porous media. As traditional methods have low  reconstruction efficiency and deep learning methods have poor generalization, a novel method for stochastic reconstruction that combines traditional methods with deep learning methods was proposed in this study. The proposed method inherited the framework of the SA method and used  the efficiency of the convolutional neural network to overcome the limitations of accuracy and speed. Moreover, this method reconstructed an arbitrary size using the SA method. Finally, the effectiveness of the proposed method was verified experimentally.

The remainder of this paper is organized as follows. Section 2 comprises the detailed motivation and implementation process of the proposed method. Section 3 describes the experimental process of the proposed method using different reference images with an analysis of the reconstruction results. The summary and prospects for future research are discussed in Section 4. Finally, the conclusions are presented in Section 5.

\section{Method}

\subsection{Motivation}

We reviewed the classical SA method, which primarily includes the state exchange process, energy function, and annealing criterion. The state-exchange process generates new states by randomly swapping two points in different phases. The energy function characterizes the information in the reference image. Statistical feature functions are often adopted because they describe statistical and structural information well. Finally, the annealing criterion determines whether the new state is accepted by calculating the change in the energy function before and after the state-exchange process.

The SA method has generality, but it also has some limitations compared with the current fast and efficient reconstruction based on deep learning methods. The SA framework is described as follows.  

\textbf{State exchange process}: The random exchange of two points is a local state exchange process in a 3D structure, and it does not involve global state exchange; this, it requires several state exchange processes. Few effective exchange points were present in the later stages of reconstruction. However, numerous exchange points are required to produce a new valid state, which requires a large number of iterative processes. In the entire reconstruction process, the energy function and annealing criterion need to be determined; therefore, the reconstruction efficiency is relatively low. When the size of the reconstruction increases, its efficiency is further reduced.

\textbf{Energy function}: The SA method can comprise different energy functions to reconstruct different characteristic structures. However, most energy functions also carry low-order statistical feature functions, such as two-point correlation and linear path functions. Other features, such as morphological features, cannot be described by low-order functions, which results in low reconstruction efficiency while using high-order feature functions. Therefore, a balance between computational efficiency and accuracy is required for the morphological features of the reconstruction results using the SA method.

\textbf{Annealing criterion}: The annealing criterion, the metropolis method, allows the energy function to converge gradually. By controlling the cooling schedule by choosing the appropriate temperature and rate of change value, the system can evolve to the target state and is not trapped by the local energy minimum. It is a practical approach, but it does not effectively guide the process of the state exchange. Accordingly, the reconstruction efficiency is reduced, and the target state cannot be reached quickly under the annealing criterion.

The review of deep learning methods mainly includes neural networks, loss functions, and gradient optimization methods. The neural network is designed to fit a mapping relationship to complete the corresponding task. The loss function describes the difference between the true value and the estimated value, and the corresponding gradient optimization method guides the updating of the neural network. Based on deep learning-based methods, mapping from the source domain to the target domain could be directly established owing to  the fitting ability of neural networks. Once the neural network containing the mapping relationship from 2D images to 3D structures is constructed, the reconstruction is a forward process, making the end-to-end reconstruction efficient.

However, most current deep learning methods depend on the 3D structure as a training sample. The information inside the reference image used by traditional methods, such as SA, was not considered. The network design and training are not based on the reference image to be reconstructed but on the training samples of the 3D structure with prior information, which leads to several fundamental problems. When no 3D structure is present as a training sample, neither training nor reconstruction can be performed. In addition, when the structural and morphological features of the reference image and the training samples are inconsistent, the neural network cannot complete reconstruction effectively even if trained, resulting in its inability to achieve generalizations, such as that in SA.

\begin{figure}[h]
	\centering
	\includegraphics[scale=.6]{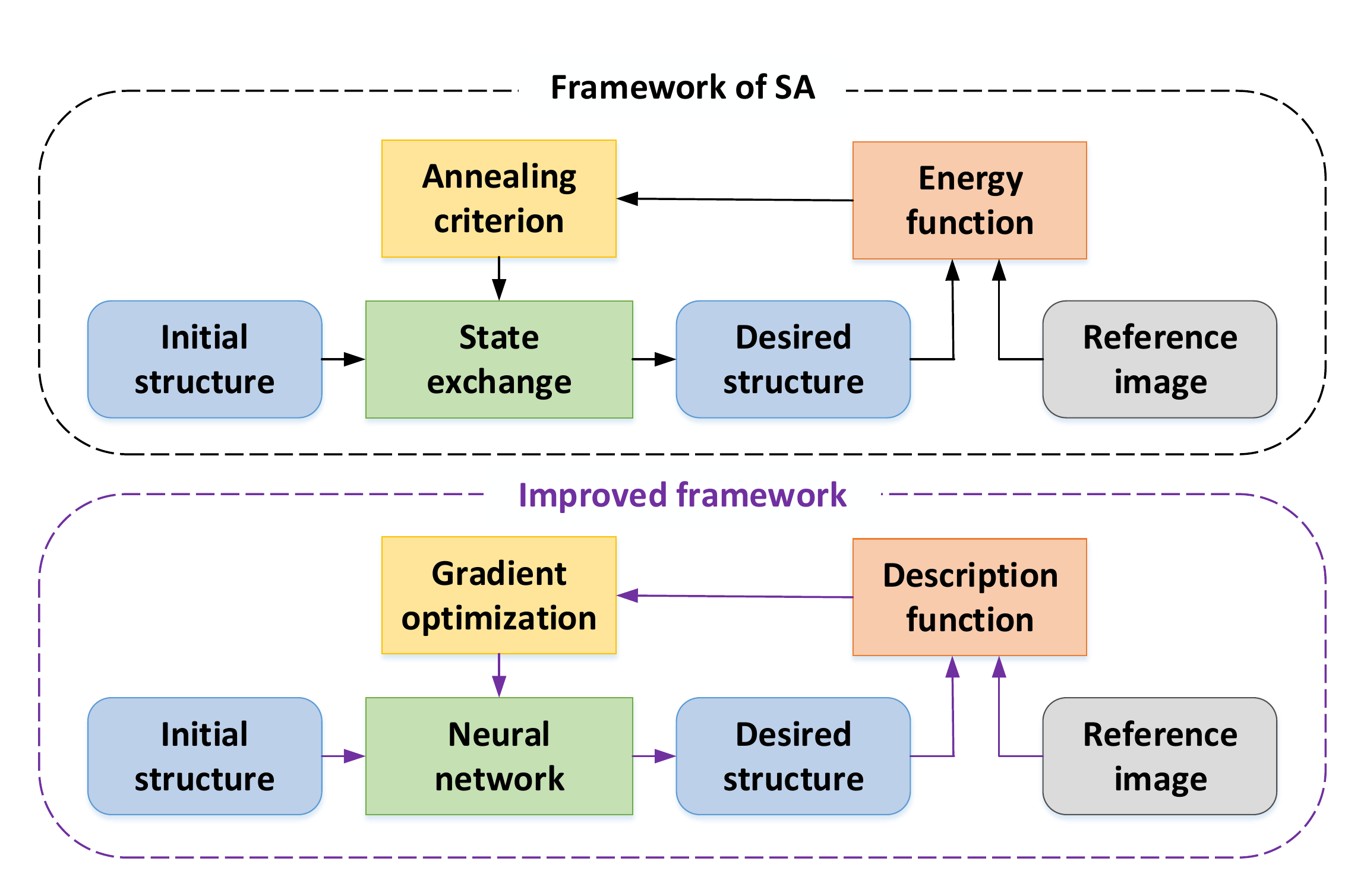}
	\caption{Simulated annealing framework (top) and improved framework of the proposed method (bottom). In the improved framework, state exchange, energy function and annealing criterion are replaced by neural network, description function and gradient optimization, respectively.}
	\label{FIG:1}
\end{figure}

\begin{figure}[h]
	\centering
	\includegraphics[scale=.4]{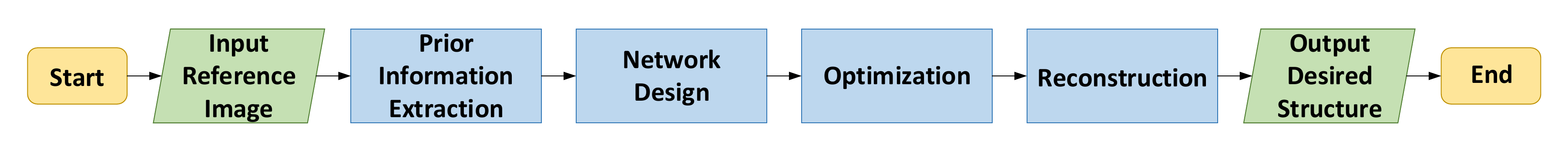}
	\caption{The flow chart of the proposed method. The flow includes four processes: prior information extraction, network design, optimization and reconstruction.}
	\label{FIG:2}
\end{figure}

Considering the above factors, we propose a reconstruction method combining the traditional SA and deep learning methods, called neural networks, with an improved simulated annealing framework (ISAF-NN). This method inherits the SA framework and combines the fast and efficient performance of neural networks in deep learning. In the proposed method, the state exchange process is replaced by the neural network, description function replaces the energy function, and gradient optimization method replaces the annealing criterion. The overall framework of the proposed method is illustrated in Figure \ref{FIG:1}. The flow of the proposed method is divided into four processes, namely prior information extraction, network design, optimization, and reconstruction, which are introduced later in this section. A flow chart of the proposed method is shown in Figure \ref{FIG:2}.

\subsection{Prior information extraction and Network design}

Prior information extraction comprises extracting structural parameters from a given 2D reference image to guide network design, which includes the neural network design and description function.

\begin{figure}[h]
	\centering
	\includegraphics[scale=.6]{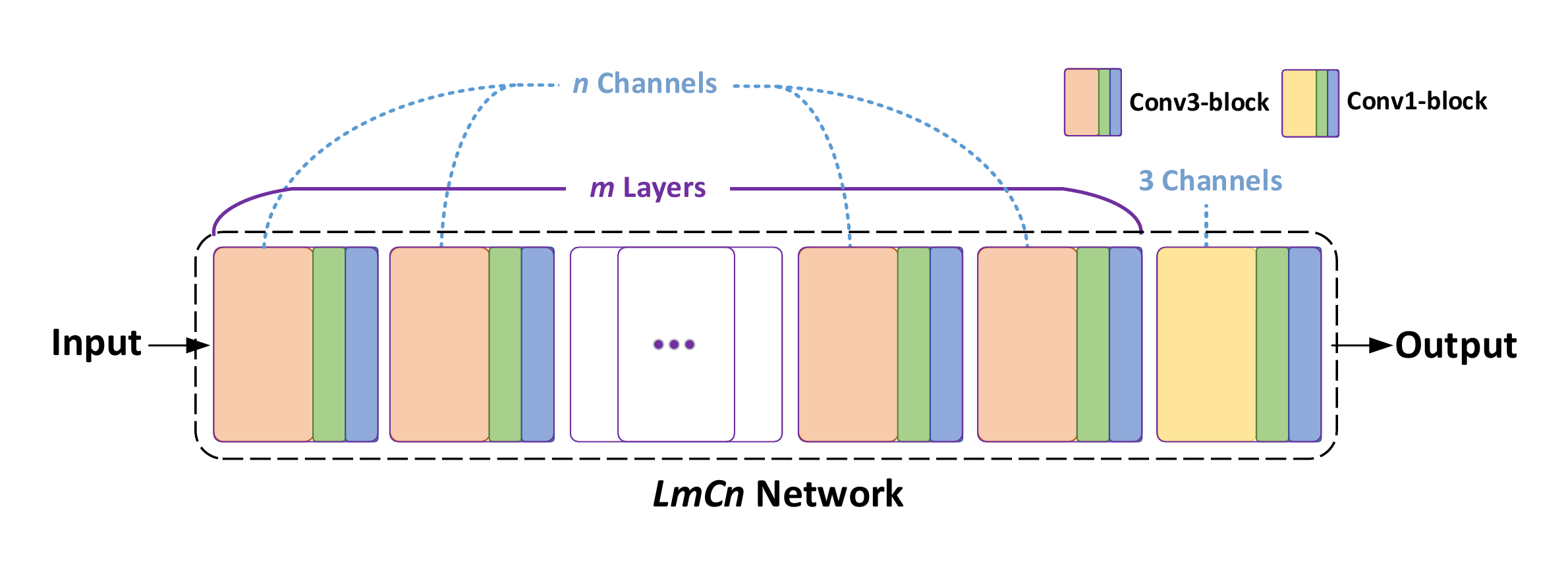}
	\caption{The structure of designed network. The network structure consists of $m$ Conv3-blocks with $n$ channels and 1 Conv1-block with 3 channels.}
	\label{FIG:4}
\end{figure}

\begin{figure}[h]
	\centering
	\includegraphics[scale=.6]{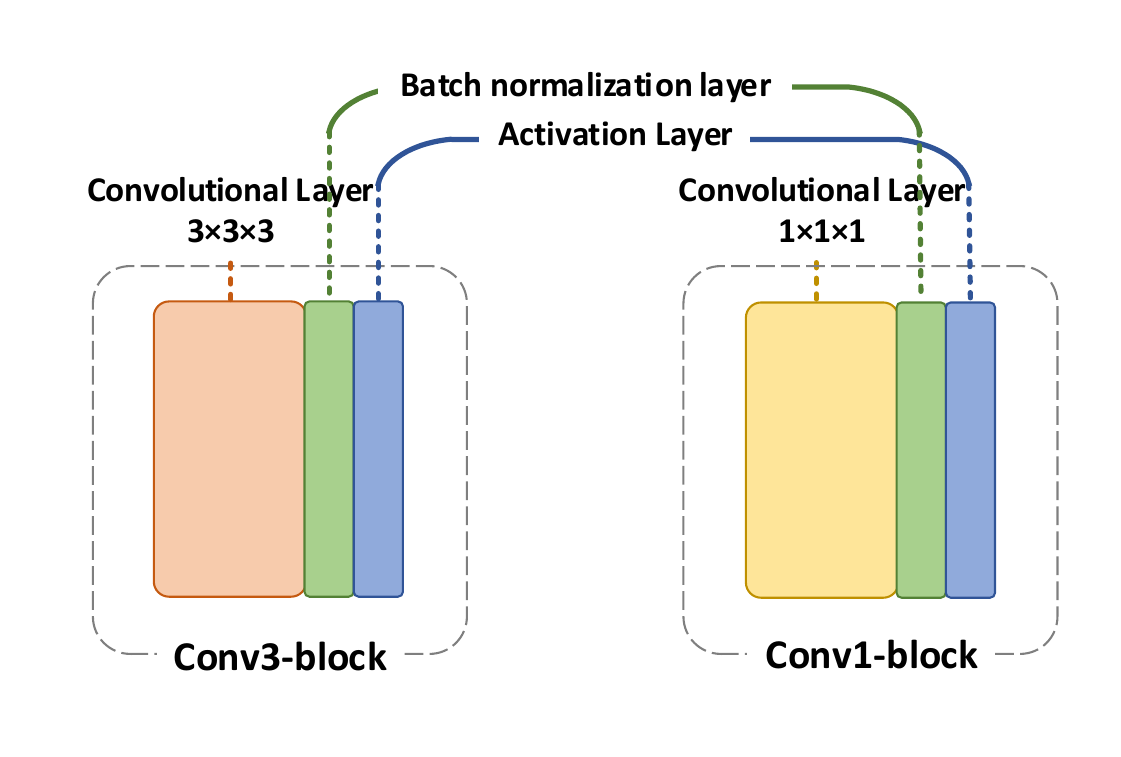}
	\caption{The basic unit structure of Conv3-block and Conv1-block. Both Conv3-block and Conv1-block consist of a convolutional layer, a normalization layer and an activation layer. The kernel size of Conv3-block and Conv1-block convolution layer is 3 and 1, respectively.}
	\label{FIG:3}
\end{figure}

As shown in Figure \ref{FIG:4}, the neural network structure is a $m$-layers and $n$-channels network ($LmCn$), which is composed of $m$ Conv3-blocks with $n$ channels and 1 Conv1-block with 3 channels. The basic structure of Conv3-block and Conv1-block is shown in the Figure \ref{FIG:3}. For Conv3-block, it consists of a 3-size 3D convolution kernel, a batch normalization (BN) layer and a Leaky Rectified Linear Unit (LeakyReLU) activation function. For Conv1-block, it consists of a 1-size 3D convolution kernel, a BN layer and a LeakyReLU activation function.

The designed $LmCn$ network aims to establish mapping by continuously updating the state similar to that of the SA and achieving  the reconstruction of 3D noise to the target 3D structure. Because  the convolutional layers of Conv3-block and Conv1-block use no padding, it maps a 3D block conforming to the noise distribution to the point of the target structure, where the size of the 3D block corresponds to the size of the receptive field of the neural network. The mapping is similar to the process of point reconstruction, in which the value of the point to be reconstructed is inferred from the surrounding data.


The size of the receptive field is of importance to the designed network. Networks with receptive fields of different sizes generate different reconstruction results. If the receptive field of the designed network is small, the reconstruction results cannot reproduce the structural features of the reference image. However, a sizeable receptive field can also lead to reduced reconstruction efficiency. Therefore, when designing the $LmCn$ network, the size of the 3D block should account for the input noise distribution, that is, the size of the receptive field of the $LmCn$ network should be proximate or slightly larger than the size of the structural parameter.

\begin{figure}[h]
	\centering
	\includegraphics[scale=.35]{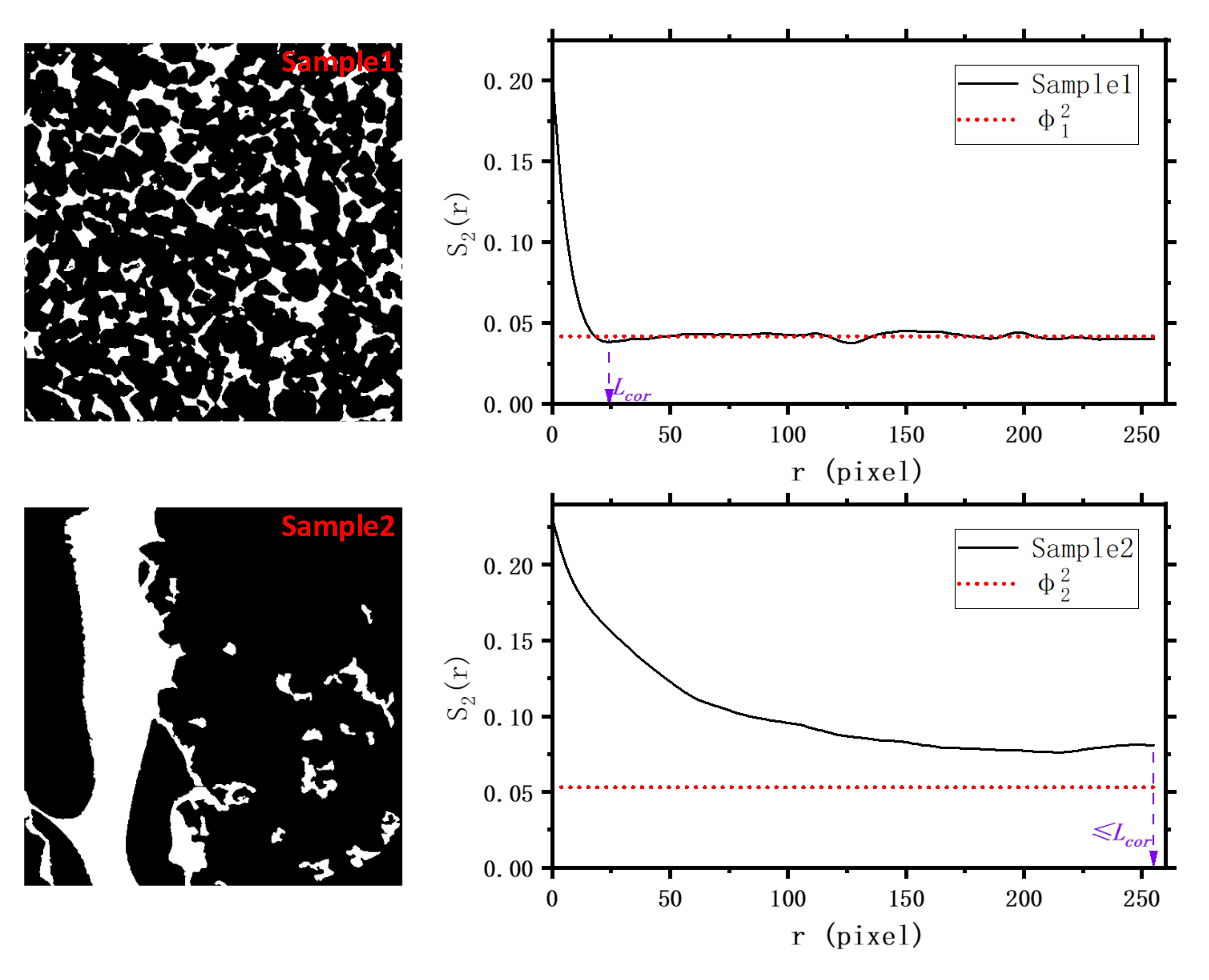}
	\caption{Schematic of two-point probability function $S_{2}(r)$ and autocorrelation distance $l_{cor}$. Sample 1 and sample 2 are homogeneous and heterogeneous images, respectively (left). The two-point probability function $S_{2}(r)$, the square of porosity $\phi^2$, and the autocorrelation distance $l_{cor}$ are shown on the right.}
	\label{FIG:6}
\end{figure}

The autocorrelation distance of the reference image was chosen as a parameter to characterize the structure of the reference image. The two-point probability function, defined as the probability that two points separated by $r$ are in the same phase, was used as an energy function for the SA method \cite{9yeong1998reconstructing}. For a homogeneous or second-order stationary structure, the function value gradually decreases as the distance increases. When a certain distance is reached, the two points are no longer correlated, and the function value fluctuates around the square of the first moment. This distance is also the autocorrelation distance, which has relatively high representation power for structural features. For a heterogeneous structure, the value of the two-point probability function does not converge but fluctuates around a specific value. In such cases, the autocorrelation distance is considered larger than the size of the heterogeneous structure image, as shown in Figure \ref{FIG:6}.

The mathematical expression for the two-point probability function is given by Equation \ref{Eq:1}. $I(x)$ is the indicator function of the pore phase in Equation \ref{Eq:2}, and $x$ is a point in an $n$-dimensional space. For homogeneous structures, the two-point probability function depends only on the relative position and not on the absolute position in space. Thus, it can be expressed using Equation \ref{Eq:3}, where $r$ is the relative position.

\begin{equation}\label{Eq:1}
	S_2 (x_1 ,x_2 ) = \left\langle {I(x_1 ),I(x_2 )} \right\rangle
\end{equation}

\begin{equation}\label{Eq:2}
	I(x) = \left\{ {\begin{array}{*{20}c}
			{1,x \in pore{\rm{ }}phase}  \\
			{0,x \notin pore{\rm{ }}phase}  \\
	\end{array}} \right.
\end{equation}

\begin{equation}\label{Eq:3}
	S_2 (r) = S_2 (x,x + r) = \left\langle {I(x),I(x + r)} \right\rangle 
\end{equation}

\begin{figure}[h]
	\centering
	\includegraphics[scale=.6]{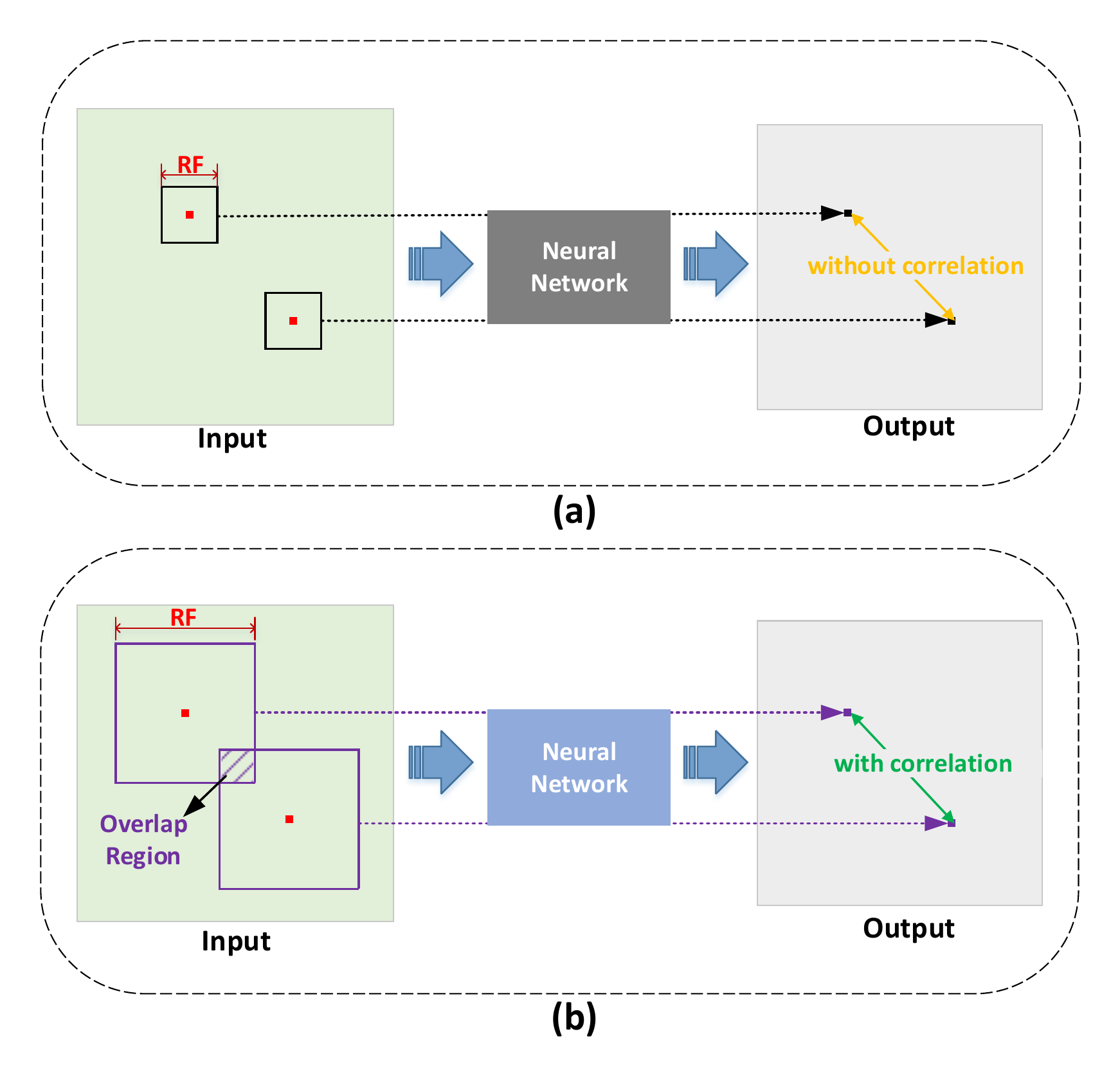}
	\caption{Schematic diagram of the correlation established for networks with different receptive fields. (a) Networks with small receptive fields. (b) Networks with large receptive fields.}
	\label{FIG:7}
\end{figure}

Consequently, the autocorrelation distance of the reference image is chosen as the guiding parameter for the $LmCn$ network design. For the input structure of the noise distribution, the autocorrelation distance is one; that is, two points with any length apart do not correlate. However, this is not the case in the reference image, where the autocorrelation distance is a certainly determined value $r$. The 3D block of the input noise distribution corresponding to the two points of the reconstruction result has a specific overlap region to reproduce the structural feature of the reference image in the reconstruction results and correlate the two points at the distance $r$ of the reconstruction results, as shown in Figure \ref{FIG:7}. The overlap region can provide specific correlation information inputs and establish the same relationship in the reconstruction results, where the minimum 3D block of the input noise distribution required is the region in the autocorrelation distance size.

To reproduce the structural features of the reference image, the size of the receptive field of the $LmCn$ network should be consistent with the autocorrelation distance or greater than the autocorrelation distance when designing the network. However, a sizeable receptive field is not recommended. An extensive receptive field increases the number of network layers and parameters, reduces the training speed, and increases the video memory consumption. Thus, it is essential to follow the size of the receptive field of the $LmCn$ network should be as close as possible to or slightly larger than the size of the structural parameter.

In general, the receptive field $l_i$ of layer $i$ is given by the Equation \ref{Eq:4}. In the equation, $f_j$ is the convolution kernel size of layer $j$ and $s_j$ is the stride of layer $j$. For the designed network $LmCn$, its receptive field can be clearly expressed by the Equation \ref{Eq:5}.

\begin{equation}\label{Eq:4}
	l_i  = l_{i - 1}  + (f_j  - 1)*\prod\limits_{j = 1}^{i - 1} {s_j }
\end{equation}

\begin{equation}\label{Eq:5}
	RF = 3 + 2*(m - 1)
\end{equation}

According to the network design criteria, the receptive field of the $LmCn$ network should be close to or slightly larger than the autocorrelation distance (Equation \ref{Eq:6}). The final number of designed network layers must satisfy the requirements of Equation \ref{Eq:7}, $L_{cor}$ is the autocorrelation distance of the reference image.

\begin{equation}\label{Eq:6}
	RF \approx L_{cor}
\end{equation}

\begin{equation}\label{Eq:7}
	m = \left\lceil {\frac{{L_{cor}  - 3}}{2}} \right\rceil  + 1
\end{equation}

For the designed $LmCn$ network, the number of layers, $m$, is mainly determined by the structural parameters of the reference image, and the number of channels, $n$, mainly considers the complexity of the distribution represented by the reference image. Complex reference image distributions lead to challenges in establishing a mapping from the source domain to the target domain by the network. For example, for two different reference images, the structural parameters may be the same, but the complexity of the distribution may differ, such as varying pattern distributions. While the other parameters remain the same, reference images with more complex pattern distributions require networks with more channels to achieve the same results. However, it is difficult to determine the minimum number of channels directly, $n$, to characterize the distribution complexity of the reference image. At the same time, the method of increasing the complexity of the mapping is to increase the number of channels $n$ and layers $m$. The number of layers $m$ is more critical than the number of channels $n$. Here, the parameter $n=16$, obtained in our experiments, was adopted. In this case, most reference images can be reconstructed by adjusting the number of layers $m$. The value of $n$ was not altered and may be further explored in subsequent studies.

The appropriate description function can be selected according to the corresponding reconstruction requirements for the design of the description function, which can be neural networks or traditional statistical functions. If the structural features of the target structure must be reproduced, traditional statistical feature functions, such as two-point probability functions, can be used. However, the two-point probability function cannot accurately describe the morphological description of the pore space. Instead, a VGG network can be adopted to reproduce the morphological characteristics of the pore space \cite{52vgg}. The VGG network is better than the two-point correlation function in the description of pore space morphology; however, it also has some limitations including the inability to accurately reproduce its structural features for regular spatial structures, which a two-point correlation function can do better. Therefore, different description functions can be performed according to the different reconstruction requirements. The VGG network was used as the description function in this study to reproduce the morphological characteristics of the pore space with good accuracy.

\subsection{Optimization}
\subsubsection{Basic framework}

The optimization process corresponds to the entire SA state exchange process. However, when the entire state-exchange process is complete, only a reconstructed 3D structure is obtained. For the optimization process, the goal was to obtain a mapping from the noise domain to the target domain. As shown in Equation \ref{Eq:8}, where $G$ is the mapping established by the $LmCn$ network, $Z$ is the noise domain that follows a uniform distribution, and $D$ is the target domain of the distribution characterized by the description function.

\begin{equation}\label{Eq:8}
	G:Z \to D
\end{equation}

During optimization, it is necessary to continuously sample from the noise domain to establish mapping $G$. From the uniformly distributed noise domain, the noise of $K$ 3D structures is sampled (Equation \ref{Eq:9}.). Each noise structure $X_k (k = 1,2,...,K)$ follows a uniform distribution and is slightly larger than the reference image due to unpadded convolutional layers.

\begin{equation}\label{Eq:9}
	X_k  \sim U(0,1),{\rm{  }}k = 1,2,...,K
\end{equation}

Subsequently, $X_k$ is fed into the $LmCn$ network to obtain $Y_k$, which is given by Equation \ref{Eq:10}. In the equation, $\theta ^{(t)}$ is the network parameter after $t$ updating iterations.

\begin{equation}\label{Eq:10}
	Y_k  = g(X_k ;\theta ^{(t)} ),{\rm{  }}k = 1,2,...,K
\end{equation}

Subsequently, a point $(x, y, z)$ is randomly chosen in the 3D structure $Y_k$, and its three sections, $Slice_{k}$, are extracted. The similarity between the three sections and the reference image was measured using a VGG description function, which is given by Equation \ref{Eq:11}. In the equation, $S_{k}^m$ and $R_k^m$ are the Gram matrices of the $m$-th layer features, which are extracted from the $Slice_{k}$ and reference image $I$, respectively. $S_{k}^m [i,j]$ is the component of the Gram matrix $S_{k}^m$ at position $(i,j)$. The calculation of Gram matrix is given by Equation \ref{Eq:12}. It is defined as the second-order cross-moment of features $F_n^m [i]$ and $F_n^m [j]]$, where $F_n^m [i]$ is the component of the vector at position $i$, flattening the $m$-th layer feature map into a $1 \times n$ vector.

\begin{equation}\label{Eq:11}
	\begin{array}{l}
		{\cal L}_{total}  = \sum\limits_k {{\cal L}_{Gram} (Slice_{k}, I)}  \\ 
		{\rm{      }} = \sum\limits_k {\sum\limits_{m = 1}^M {\frac{1}{{N_m^2 }}} \left\| {S_{k}^m  - R_k^m } \right\|^2 }  \\ 
	\end{array}
\end{equation}

\begin{equation}\label{Eq:12}
	\begin{gathered}
		S_k^m [i,j] = \mathbb{E}[F_n^m [i]F_n^m [j]] \hfill \\
		{\text{     }} = \frac{1}
		{{N^m }}\sum\limits_n {F_n^m [i]F_n^m [j]}  \hfill \\ 
	\end{gathered} 
\end{equation}

Finally, the obtained loss is backpropagated to the $LmCn$ network through the gradient to update the network parameters according to Equation \ref{Eq:13}. In the equation, $\theta ^{(t)}$ is the $LmCn$ network parameters after $t$ iterations, $\hat u^{(t)}$ is the corrected value of the gradient mean $u^{(t)}$ in the $t$-th iteration, $\hat v^{(t)}$ is the corrected value of the gradient variance $v^{(t)}$ in the $t$-th iteration, $\alpha$ is the learning rate and $\varepsilon$ is a smoothing term. It is necessary to correct the bias of the gradient mean $u^{(t)}$ and variance $v^{(t)}$ to reduce the impact of bias on the initial training stage. The expressions for the corrected values $u^{(t)}$ and $v^{(t)}$ are given by Equations \ref{Eq:14}–\ref{Eq:17}. The Adam method \cite{532014Adam} was used for gradient optimization. In Adam method, the parameter $\beta _1$ and learning rate $\alpha$ are set to 0.1, the parameter $\beta _2$ is set to 0.999, and $\varepsilon$ is set to $10^{-8}$. The pseudocode for the optimization process is shown in Algorithm \ref{alg:Opt}.

\begin{equation}\label{Eq:13}
	\theta ^{(t)}  = \theta ^{(t - 1)}  - \alpha \frac{{\hat u^{(t)} }}{{\sqrt {\hat v^{(t)} }  + \varepsilon }}
\end{equation}

\begin{equation}\label{Eq:14}
	u^{(t)}  = \beta _1 u^{(t - 1)}  + (1 - \beta _1 )d\theta ^{(t)2} 
\end{equation}

\begin{equation}\label{Eq:15}
	\hat u^{(t)}  = \frac{{u^{(t)} }}{{1 - \beta _2^t }}
\end{equation}

\begin{equation}\label{Eq:16}
	v^{(t)}  = \beta _2 v^{(t - 1)}  + (1 - \beta _2 )d\theta ^{(t)} 
\end{equation}

\begin{equation}\label{Eq:17}
	\hat v^{(t)}  = \frac{{v^{(t)} }}{{1 - \beta _1^t }}
\end{equation}

\begin{algorithm}[h]  
	\caption{Optimization}  
	\label{alg:Opt}  
	\begin{algorithmic}[1]  
		\Require  
		Reference image $I$; Number of iterations $T$; Batch size $K$;
		\Ensure 
		Parameters of $LmCn$ networks $\theta$;
		\\Let $t \leftarrow 0$,initialize $\theta$
		\Repeat 
		\For{k=1 to K}
		\State Generate a 3D noise $X_k  \sim U(0,1)$,and generate 3D output $Y_k  = g(X_k ;\theta ^{(t)} )$.
		\State Sample a point randomly and extract the corresponding sections $Slice_k$ of the 3D output $Y_k$.
		\State Compute the description function loss according to Equation \ref{Eq:11} using the reference image $I$ and the extracted sections.
		\EndFor
		\State Update $\theta ^{(t)}$ according to Equation \ref{Eq:13}.
		\\Let $t \leftarrow t + 1$
		\Until{$t=T$}
	\end{algorithmic}  
\end{algorithm}

\subsubsection{Improved framework combining SA}

Combining the SA method and reconsidering the above optimization process to optimize the above framework, the entire process can be realized by initializing a noisy 3D structure, such as SA.

Our goal is to establish a mapping of the noise domain to the target domain. For any point in the target domain, the three sections corresponding to the point are morphologically and structurally similar to the reference image. When using the description function to measure similarity, the approach is to find a random point in the output 3D structure and then extract its three sections. Regardless of how large the noise structure is initialized, or even if the infinite noise domain is fed into the $LmCn$ network, the useful information used in the description function is only the information of three sections corresponding sections to the final sampling point. Only the corresponding noise region of the three sections must be input to achieve the same effect and speedup.

Furthermore, it is necessary to ensure that there is enough sampling data for the model to establish such a mapping. For a uniformly distributed noise domain, no correlation existed between the two non-overlapping sampling data points at different locations. In the entire optimization process, when initializing the noise 3D structure size is much larger than the product of the batch size and iterations, it is also sufficient to initialize the noise 3D structure sample point data. Taking a 100-size reference image as an example, the number of sampling points is one million for a noisy 3D structure initialized to the same size. However, for the relevant parameters, the batch size was set to 1, and the number of iterations was set to 1000. The product of the batch size and iterations was much smaller than the number of points sampled from the noise structure.

\begin{figure}[h]
	\centering
	\includegraphics[scale=.35]{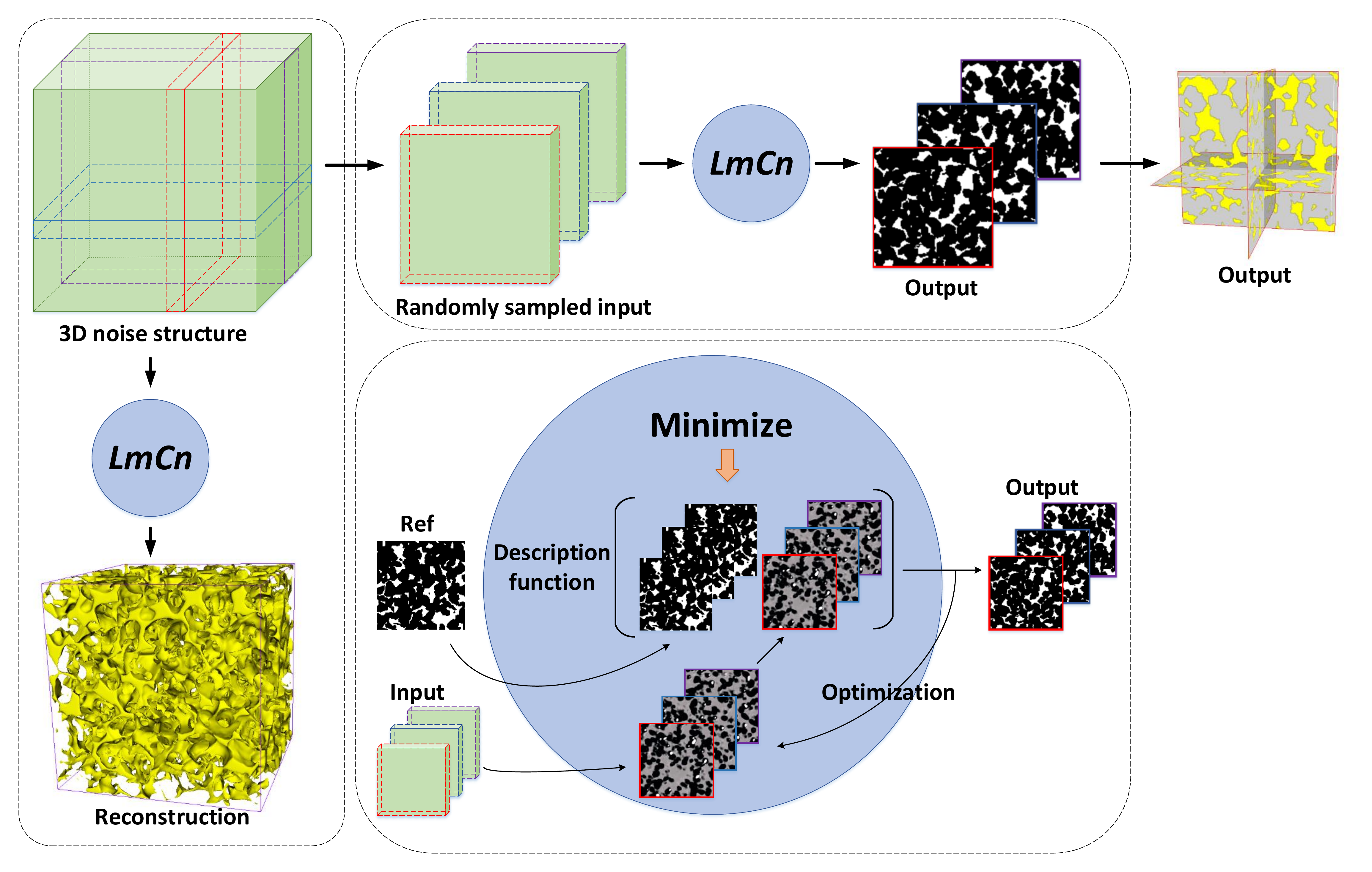}
	\caption{The explanation of optimization process. The basic optimization framework (left), the improved optimization framework (top right) and the specific process of optimization (bottom right). Where $LmCn$ is the designed network.}
	\label{FIG:8}
\end{figure}

Therefore, when Equation \ref{Eq:18} is satisfied, it is sufficient to initialize a 3D noise structure, and only the noise regions corresponding to the three sections of the final sampling point are input each time to save the video memory and accelerate the optimization process. In Equation \ref{Eq:18}, $H$, $W$ and $L$ are the length, width, and height of the initial 3D noise structure, respectively; $K$ is the batch size, and $T$ is the number of iterations. The pseudocode for the improved optimization process is presented in Algorithm \ref{alg:ImOpt}. The optimization process and improved optimization are shown in Figure \ref{FIG:8}.

\begin{equation}\label{Eq:18}
	H*W*L \gg K*T
\end{equation}

\begin{algorithm}[h]  
	\caption{Improved Optimization}  
	\label{alg:ImOpt}  
	\begin{algorithmic}[1]  
		\Require  
		Reference image $I$; Number of iterations $T$; Batch size $K$;
		\Ensure 
		Parameters of $LmCn$ networks $\theta$;
		\\Let $t \leftarrow 0$,initialize $\theta$
		\\Generate a 3D noise $X_0 \sim U(0,1)$
		\Repeat 
		\For{k=1 to K}
		\State Sample a point randomly and extract the corresponding three input regions $Input_k$.
		\State Generate the output of three sections $Slice_k$, according to $Slice_k  = g(Input_k ;\theta ^{(t)} )$.
		\State Compute the description function loss according to Equation \ref{Eq:11} using the reference image $I$ and the output sections.
		\EndFor
		\State Update $\theta ^{(t)}$ according to Equation \ref{Eq:13}.
		\\Let $t \leftarrow t + 1$
		\Until{$t=T$}
	\end{algorithmic}  
\end{algorithm}

\subsection{Reconstruction}

In contrast to the optimization process, the reconstruction process only performs a forward process because such mapping is established during the optimization process. In the reconstruction process, only the length, width, and height of the reconstruction result must be input, and the corresponding noise 3D structure is initialized and sent to the $LmCn$ network to obtain the corresponding reconstruction result. This network can reconstruct 3D structures of any size because the designed network is a fully convolutional network applicable to inputs of any size. Any 3D structure with the same size as the reference image extracted from the reconstructed structure can maintain a description function consistent with the corresponding 3D structure of the reference image. However, owing to the limitations of computer memory, it may not be possible to directly input the network to obtain the corresponding reconstruction results when reconstructing a large structure.

\begin{figure}[h]
	\centering
	\includegraphics[scale=.35]{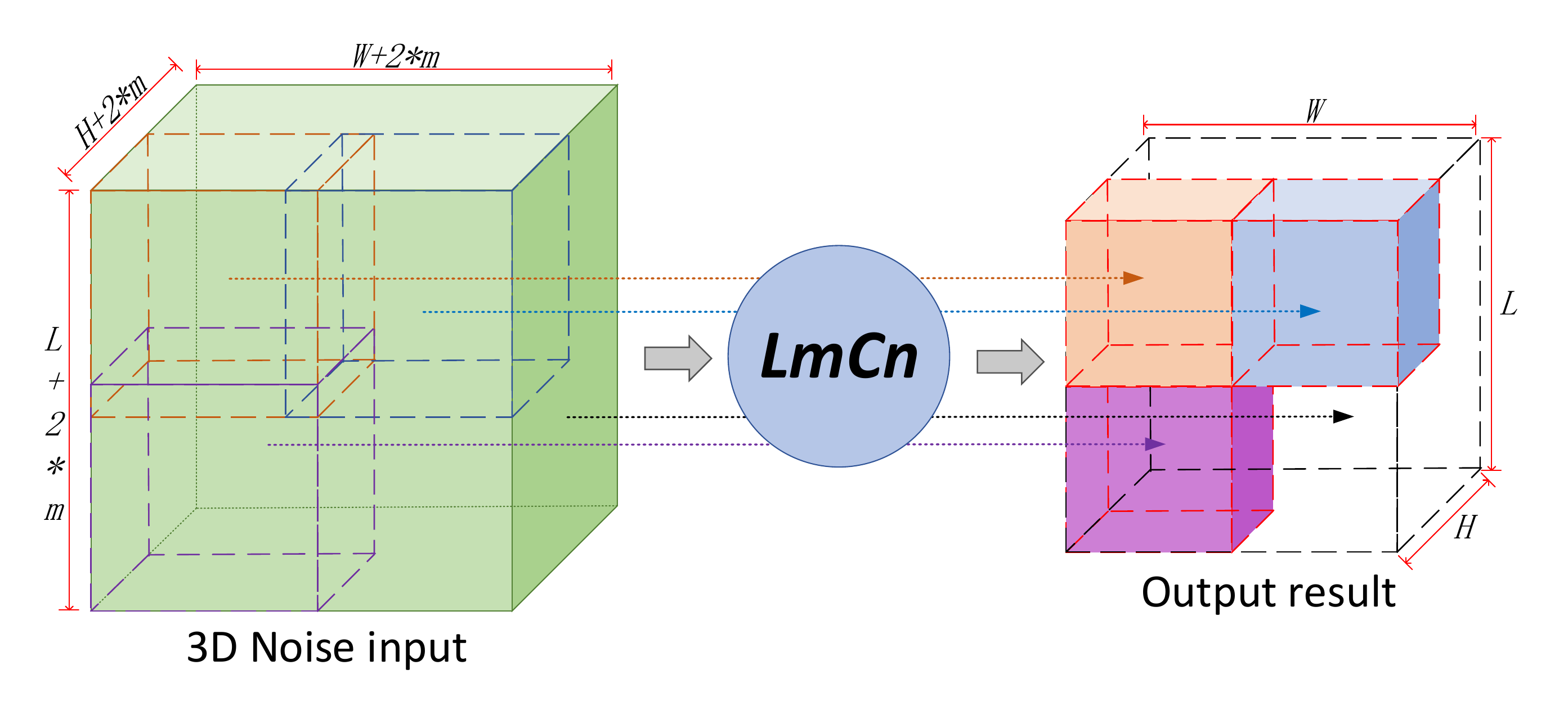}
	\caption{The explanation of reconstruction method of cutting into sub-blocks. Where $LmCn$ is the designed network.}
	\label{FIG:9}
\end{figure}

Cutting the input into small sub-blocks can be used to reconstruct the 3D structure as much as possible with limited computer memory. First, a 3D noise structure and an empty output result were initialized. Subsequently, a sub-block was cut from the noise structure and fed into the network in a specific order, and the corresponding reconstruction result is placed on the output structure. This process was repeated until the output structure was updated (Figure \ref{FIG:9}).

This method does not consider the discontinuity problem of the boundaries where the reconstruction results are concatenated. Because an unpadded convolutional $LmCn$ network is adopted, each point of the reconstruction result can find its corresponding subblock in the input noise structure. In the extreme case, it is even possible to input the noise sub-block to ensure that only one point is generated at a time and then assign the value of this point to the corresponding position of the empty reconstruction result. Thus, the largest 3D structure could be reconstructed using the least memory. However, one point is reconstructed each time, and the corresponding reconstruction times increase sharply, decreasing reconstruction efficiency. 

Therefore, when the memory is insufficient to input the entire noise 3D structure directly, the method of cutting into small sub-blocks can be used; however, a balance must be made between the number of sub-blocks and the time of reconstruction. The pseudocode for the reconstruction process is shown in Algorithm \ref{alg:Rec}.

\begin{algorithm}[h]  
	\caption{Reconstruction}  
	\label{alg:Rec}  
	\begin{algorithmic}[1]  
		\Require  
		Parameters of $LmCn$ networks $\theta$;Reconstruction size $L,H,W$;Sub-block size $l,h,w$;
		\Ensure 
		3D reconstruction result $Y$;		
		\\Load parameters of networks $\theta , m$.
		\\Initialize a 3D noise structure $X$ of size $(L+2*m)(H+2*m)(W+2*m)$, and $X \sim U(0,1)$.
		\\Initialize an empty 3D structure $Y$ of size $L*H*W$.
		\For{k=1 to L}
		\For{i=1 to H}
		\For{j=1 to W}
		\State Get the 3D sub-block $Z=X[k:k+l+2*m,i:i+h+2*m,j:j+w+2*m]$.
		\State Generate 3D reconstruction result $g(Z ;\theta )$ and update the value of corresponding region in $Y$.
		\State Let $j \leftarrow j + w$.
		\EndFor
		\State Let $i \leftarrow i + h$.
		\EndFor
		\State Let $k \leftarrow k + l$.		
		\EndFor
	\end{algorithmic}  
\end{algorithm}

\section{Experiments and results}

The experiments in this section included three parts to verify the effectiveness of the proposed method. First, the depth and width of the $LmCn$ network were explored, and reconstruction experiments were conducted on the same reference image with different network layers and channels to study the corresponding effect on the reconstruction results and verify the rationality of the proposed method. Second, to verify the effectiveness and generalization of the proposed method, four types of homogeneous porous media and an anisotropic porous medium were used as experimental materials for the reconstruction experiments. Finally, the 3D structural reconstruction of a larger size was verified. The materials used in the previous effectiveness and generalization verification experiments were used to reconstruct a 512-size 3D structure.

\subsection{Experiments for depth and width of network}

When designing the network, the receptive field of the network is similar to or slightly larger than the structural parameters of the reference image. Not only does the $LmCn$ network need to achieve the receptive field size but the description function should also be the same if the network is adopted as the description function. The features of the reference image extracted using the description function also represent the target domain. The optimization objective is unclear when the reference image cannot be effectively characterized or described. Establishing mapping and reproducing the same features as the target structure is thus complex.

In the optimization process, the VGG network is adopted as the description function, which can accurately and effectively  characterize the morphology of the pore space. Simultaneously, the receptive field of the entire VGG network is approximately 212, which is sufficient for most reference images. Detailed features are extracted in the first few layers of the VGG network, and structural features are extracted in the subsequent layers. Therefore, in the designed $LmCn$ network, reaching a specific receptive field size is necessary to reproduce the same features of the reference image.

A balance between accuracy and speed should also be considered. Too small a receptive field or insufficient network depths will fail to reproduce long-range structures. However, too deep a network leads to an increase in  parameters and a reduction in efficiency. This finding is in line with the principle of network design.

The complexity of the network intuitively represents the width of the $LmCn$ network, that is, the complexity of the mapping from this noise domain to the target domain. Because the noise domain is always subject to a fixed uniform noise distribution, the target domain is determined using the description function of the reference image. Different types of reference images correspond to different levels of target-domain complexity. Therefore, the width of the $LmCn$ network is primarily related to the complexity of the description of the reference image. A narrow network, such as a width network, cannot establish such a mapping. Similarly, a more complex reference image would require a broader network. 

\begin{figure}[h]
	\centering
	\includegraphics[scale=.35]{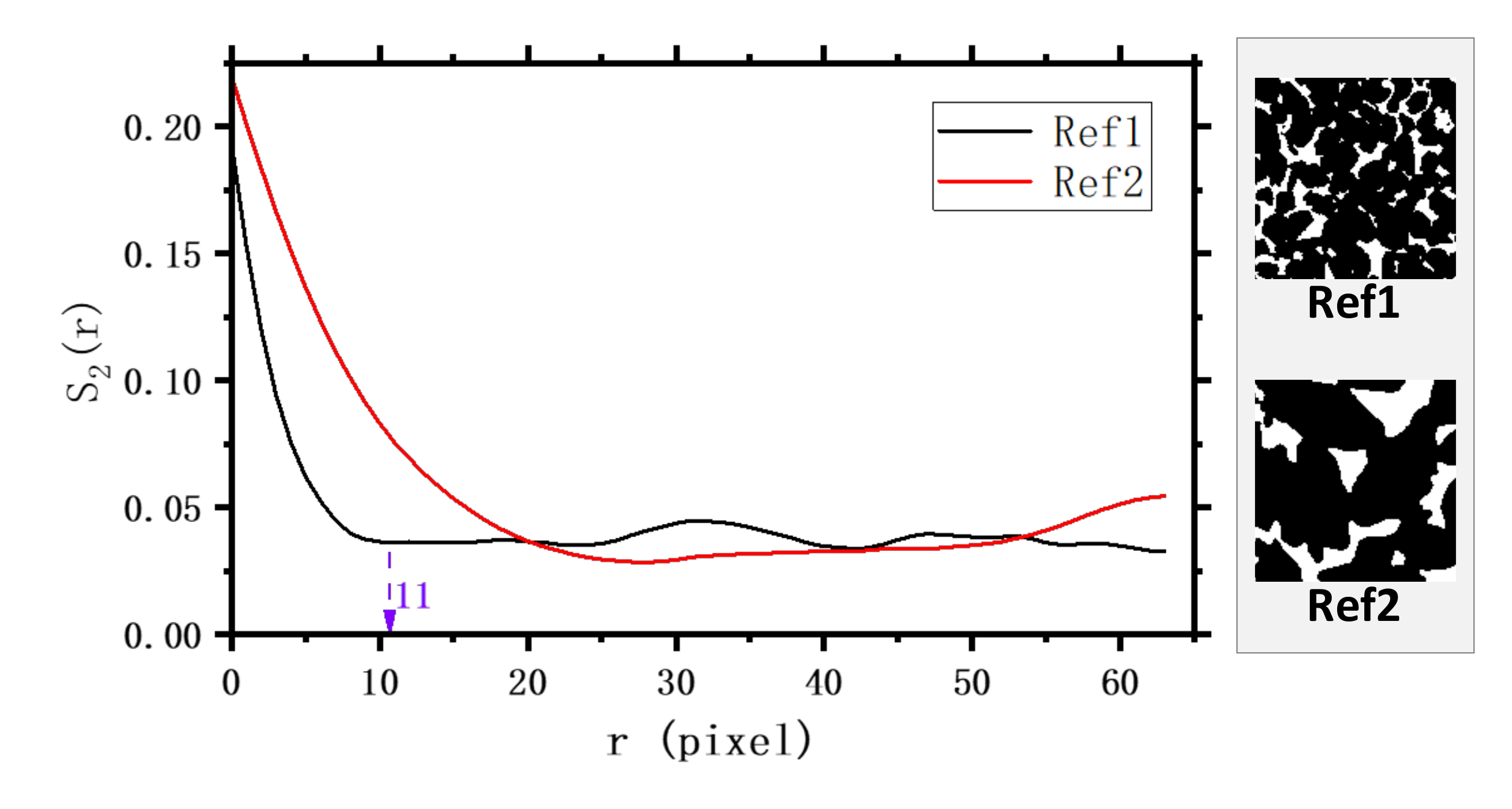}
	\caption{Two-point probability functions $S_2(r)$ of Ref1 and Ref2. The autocorrelation distance $l_{cor}$ of Ref1 is approximately 11; Ref2 shows some non-stationary characteristics, so the autocorrelation distance is not marked.}
	\label{FIG:10}
\end{figure}

Here, the effect of the depth and width of the network on the reconstruction results is also explored. In the experiment, a homogeneous 128-size sandstone image and a 128-size local area image cut from a homogeneous sandstone image of 1000 size are used for reconstruction (Figure \ref{FIG:10}).

\subsubsection{Experiments on Ref1}

A two-point probability function was calculated for the homogeneous 128-size sandstone image Ref1, and the autocorrelation distance was approximately 11. According to Equation \ref{Eq:7}, five Conv3-block layers were present. Thus, network structures of five layers with eight channels  ($L5C8$), 16 channels ($L5C16$), and 24 channels ($L5C24$) were designed and used. The designed networks were used to train the model according to the optimization process. The corresponding models were obtained by setting the learning rate as 0.1, batch size as 1, and the number of iterations as 1000. The model was then used to complete the 3D reconstruction through the reconstruction process, and a set of 3D structures of the same size as the reference image were reconstructed for verification analysis.

\begin{figure}[h]
	\centering
	\includegraphics[scale=.15]{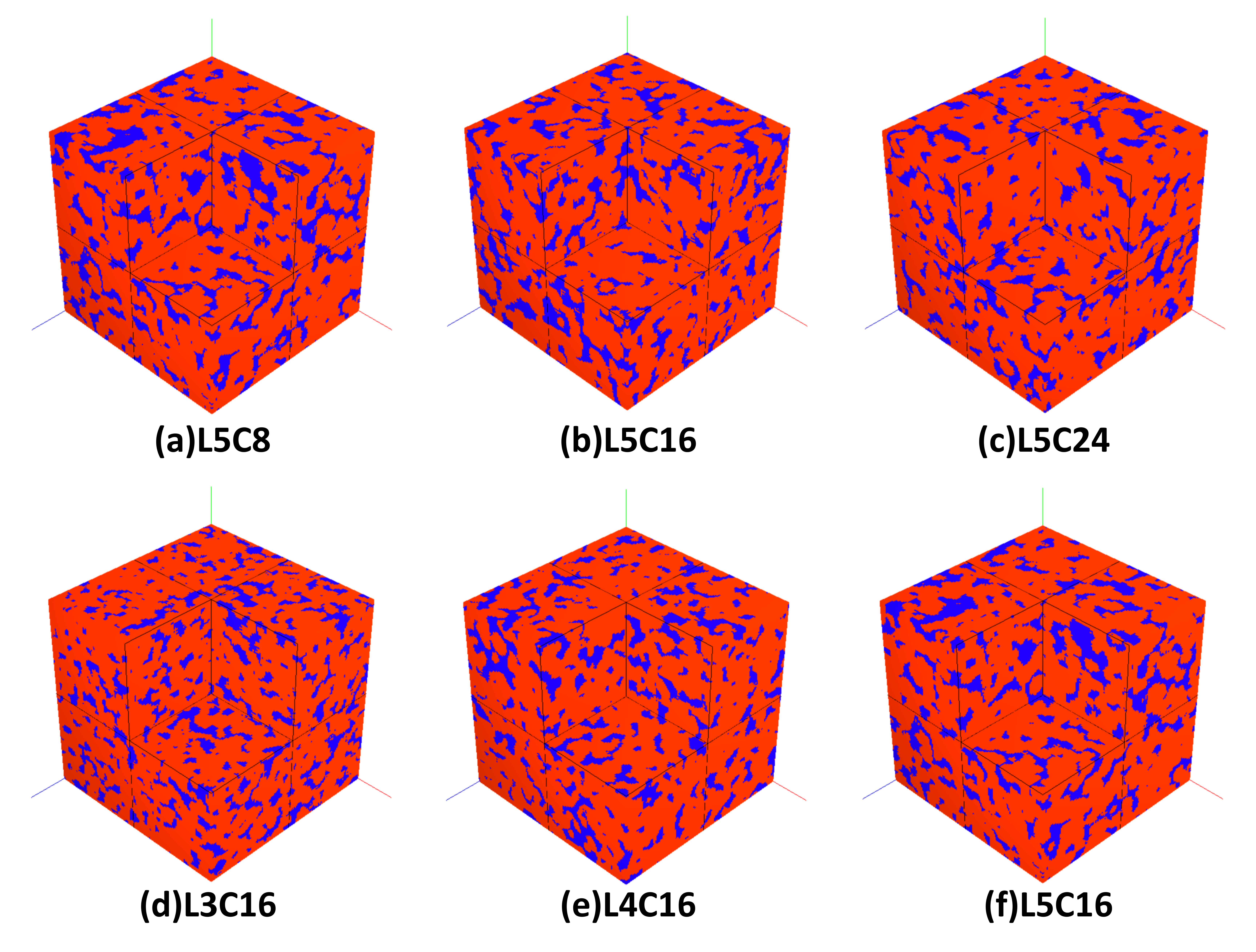}
	\caption{3D Visualization of reconstruction results using networks with different width and depth on Ref1. (a) Reconstructed structure of $L5C8$ network. (b) Reconstructed structure of $L5C16$ network. (c) Reconstructed structure of $L5C24$ network. (d) Reconstructed structure of $L3C16$ network. (e) Reconstructed structure of $L4C16$ network. (f) Reconstructed structure of $L5C16$ network.}
	\label{FIG:11}
\end{figure}

\begin{figure}[h]
	\centering
	\includegraphics[scale=.35]{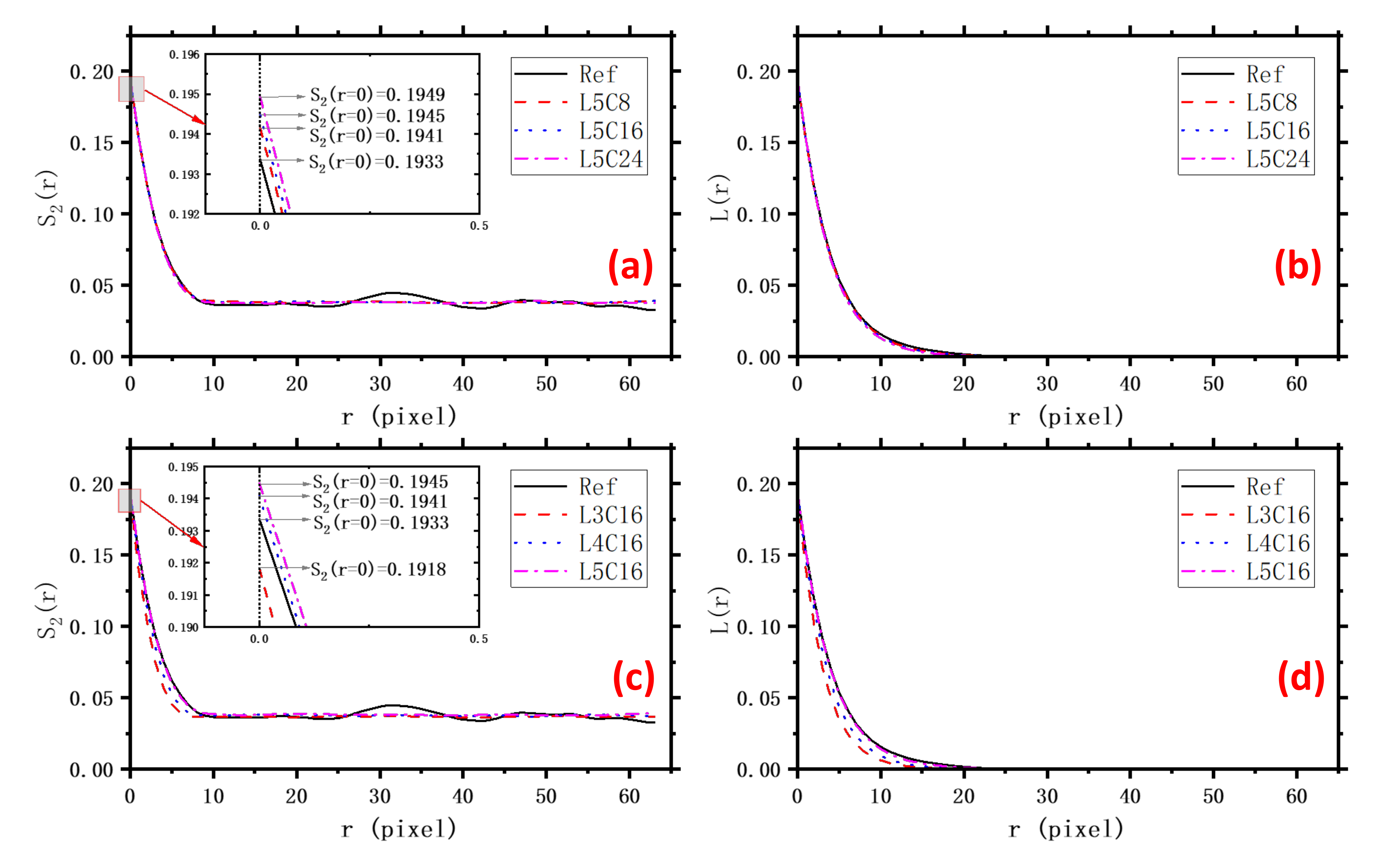}
	\caption{Comparison of statistical functions of reconstruction results using networks with different width and depth on Ref1. (a) Two-point probability functions $S_2(r)$ of reconstructed structures for different network widths. (b) Linear path function $L(r)$ of reconstructed structures for different network widths. (c) Two-point probability functions $S_2(r)$ of reconstructed structures for different network depths. (d) Linear path function $L(r)$ of reconstructed structures for different network depths.}
	\label{FIG:12}
\end{figure}

Figure \ref{FIG:11} (a)–(c) shows the reconstruction results of the networks with different widths under the same parameters. Each reconstructed 3D structure has a small piece cut out at the vertex to facilitate viewing the internal structure. Reconstruction results were visualized using the FDMSS 3D display program \cite{54gerke2018finite}. The two-point correlation function and  linear path function were calculated for evaluation to observe the relevant information of the results further (Figure \ref{FIG:12} (a)–(b)). In addition, network structures of three layers with 16 channels ($L3C16$), four layers with 16 channels ($L4C16$), and five layers with 16 channels ($L5C16$) were used to explore the impact of network depth. As with the previous setting of the network width, this set of experiments differed only in depth, with the remaining parameters consistent, and the visualization results are shown in Figure \ref{FIG:11} (d)–(f). The statistical functions are shown in Figures \ref{FIG:12} (c) and (d).

As shown in Figure \ref{FIG:11} (a)–(c) and Figure \ref{FIG:12} (a)–(b), when the network depth reaches the structural parameter requirements, the number of channels reaches a specific number, and then increasing the number of channels does not affect the reconstruction results. The reconstruction results of the three groups of channels did not differ significantly in terms of pore size, morphology, and statistical function. However, as shown in Figure \ref{FIG:11} (d)–(f) and Figure \ref{FIG:12} (c)–(d), when the number of channels is the same, changing the depth of the network will affect the reconstruction result. The pore size of $L5C16$ is larger than those of $L4C16$ and $L3C16$. This result is also observed in the statistical function of the difference in the autocorrelation distance and linear path distribution.

The larger the channels, the more complex the mapping relationships that can be fitted. According to the visualization results in Figure \ref{FIG:11} and the relevant statistical function analysis in Figure \ref{FIG:12}, the number of channels has little effect after the network reaches a certain number of layers. However, to adapt to more types of reference images, the number of channels $n=16$ was adopted in this study. If the width is too small, some reference images with complex structures may not be reconstructed; if the width is too large, it will increase memory consumption and reduce reconstruction efficiency. There was a compromise between the accuracy and efficiency in this study. 

\subsubsection{Experiments on Ref2}

For the 128-size local area image Ref2, as it exhibits some non-stationary conditions (Figure \ref{FIG:10}), it is difficult to judge its autocorrelation distance. Therefore, network structures of three layers with 16 channels ($L3C16$), five layers with 16 channels ($L5C16$), eight layers with 16 channels ($L8C16$), and 11 layers with 16 channels ($L11C16$) were designed for the reconstruction.

\begin{figure}[h]
	\centering
	\includegraphics[scale=.15]{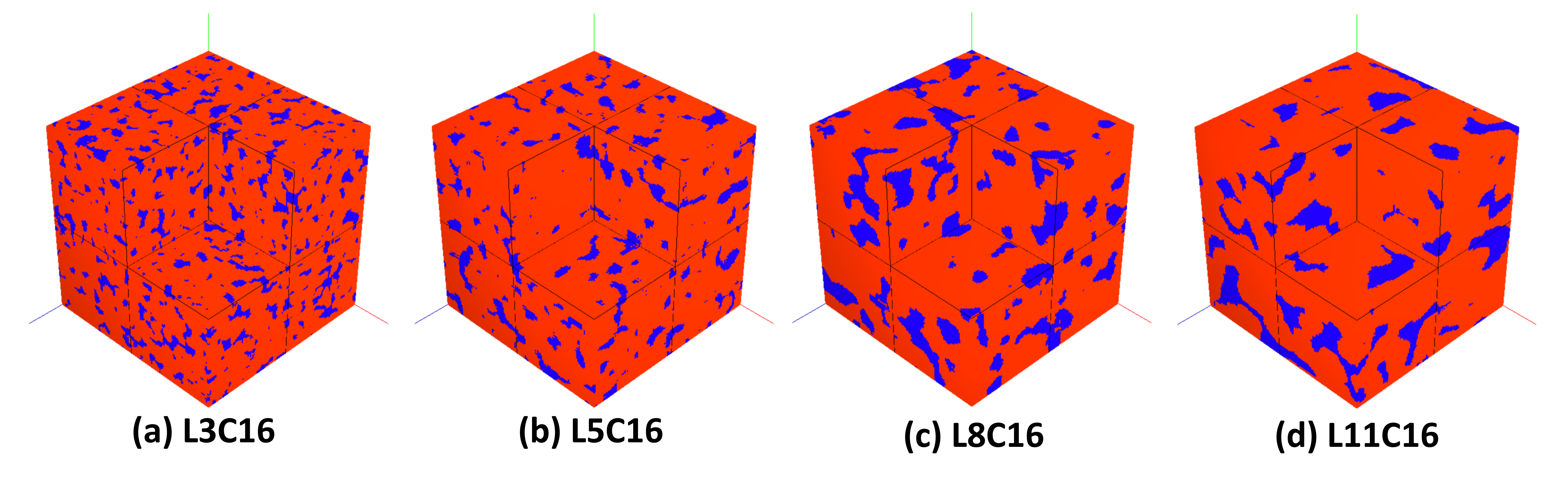}
	\caption{3D Visualization of reconstruction results using networks with different depth on Ref2. (a) Reconstructed structure of $L3C16$ network. (b) Reconstructed structure of $L5C16$ network. (c) Reconstructed structure of $L8C16$ network. (d) Reconstructed structure of $L11C16$ network.}
	\label{FIG:13}
\end{figure}

\begin{figure}[h]
	\centering
	\includegraphics[scale=.35]{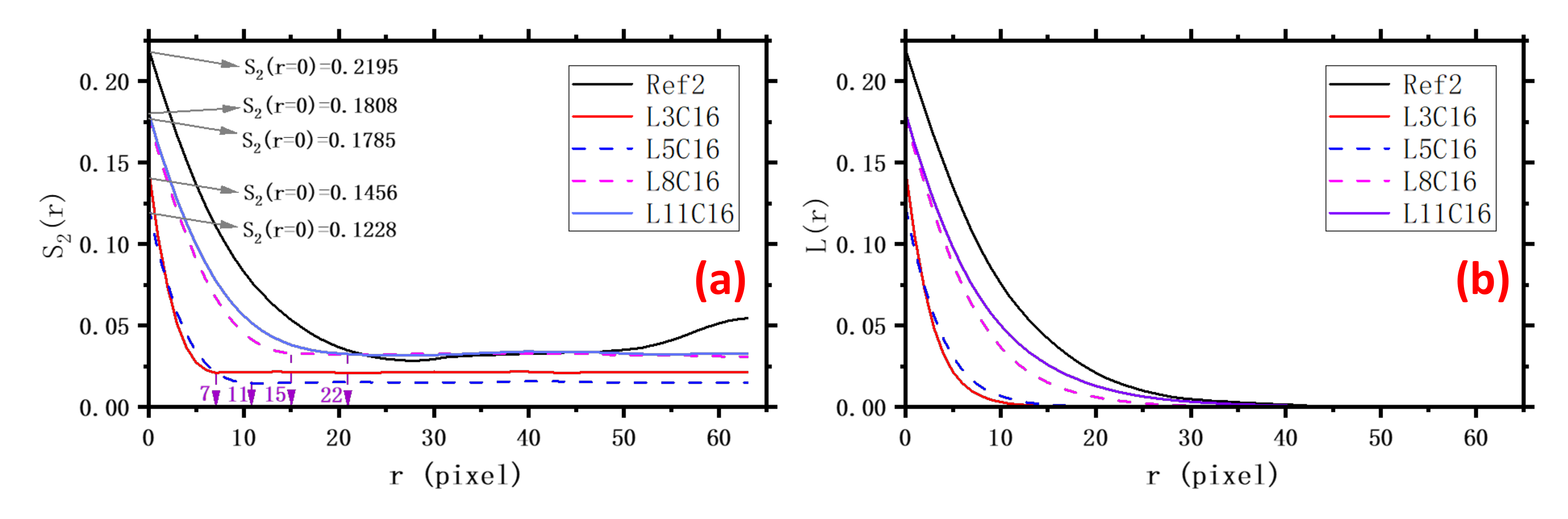}
	\caption{Comparison of statistical functions of reconstruction results using networks with different depth on Ref2. (a) Two-point probability functions $S_2(r)$ of reconstructed structures for different network depths. (b) Linear path function $L(r)$ of reconstructed structures for different network depths.}
	\label{FIG:14}
\end{figure}

The parameters used in this set of experiments were also consistent with those on Ref1, with only the number of network layers being changed. The visualization results are shown in Figure \ref{FIG:13} (a)–(d). The statistical functions are shown in Figure \ref{FIG:14} (a) and (b). From Figure \ref{FIG:13} (a)–(d), with the increase in network depth, the structural information that can be captured and established by the network becomes increasingly larger, which is observed as a change in the pore size. The deeper the network, the closer the reconstructed image's pore size is to the reference image's pore size (Figure \ref{FIG:14} (a) and (b)). The autocorrelation distance gradually increases with the increase in network depth. The linear path distribution also gradually shifts to the right to a larger pore space structure. The reconstruction results of $L11C16$ show that the autocorrelation distance is close to 22. In contrast, the autocorrelation distances of $L8C16$, $L5C16$, and $L3C16$ were also close to 17, 11, and 7, respectively. This finding is consistent with the relationship between the number of network layers m and the autocorrelation distance derived in Equation \ref{Eq:7}. Furthermore, this result also indicates that a specific receptive field size is required to reproduce the structural features of the reference image.

\subsection{Validation experiments of effectiveness and generalization}

\begin{figure}[h]
	\centering
	\includegraphics[scale=.5]{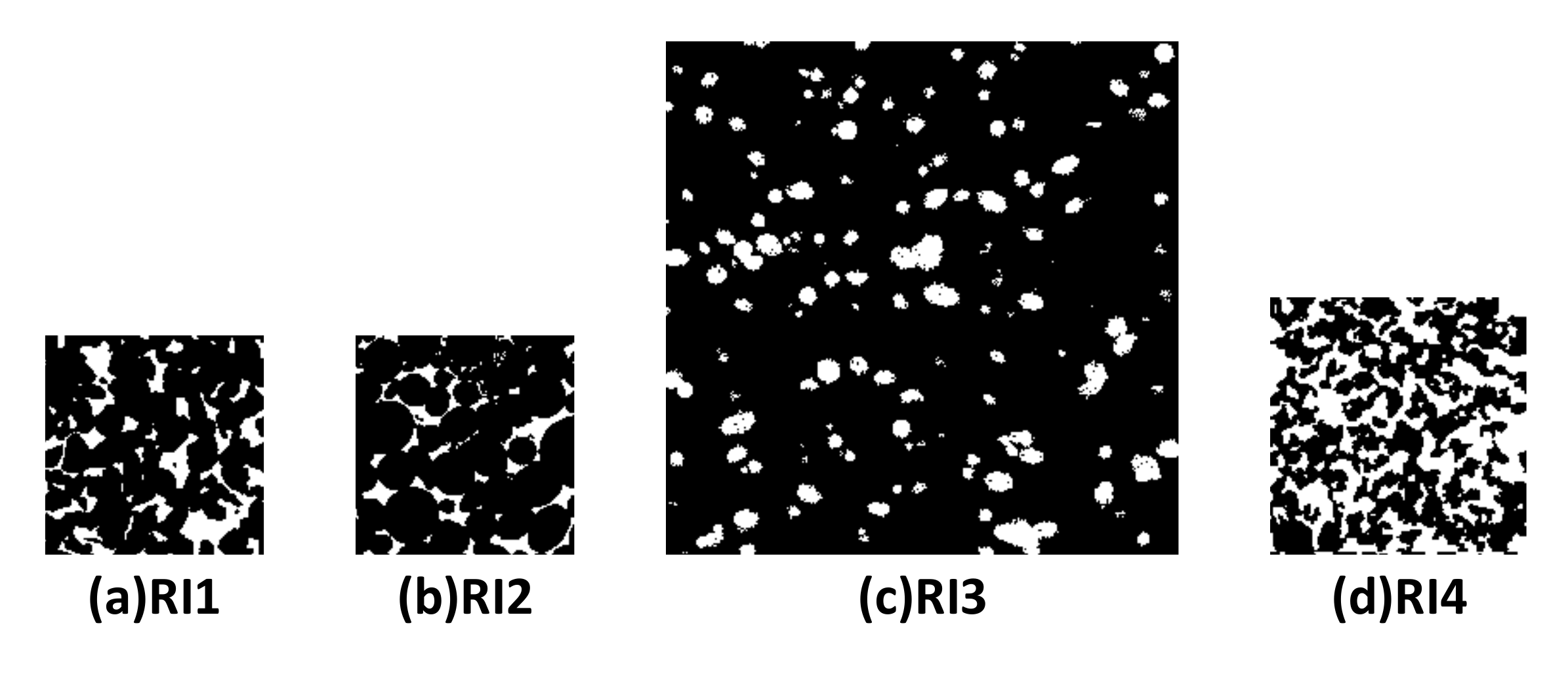}
	\caption{Different reference images of experimental materials. (a) Sandstone image RI1. (b) Limestone image RI2. (c) Silica material image RI3. (d) Synthetic ceramic material image RI4.}
	\label{FIG:15}
\end{figure}

Experiments were conducted on four types of porous media to verify the effectiveness of the proposed method. The four materials are Fontainebleau sandstone, Ketton limestone, silica, and synthetic ceramic \cite{55coker1996morphology,56dong2009pore,57muljadi2016impact}. They differ in their porosity, pore space morphology, and other parameters. Reference images of the four materials are shown in Figure \ref{FIG:15}. The 2D reference images of Fontainebleau sandstone RI1 and Ketton limestone RI2 are 128$\times$128 in size, silica material RI3 is 300$\times$300 in size, and synthetic ceramic material RI4 is 150$\times$150 in size. The point size of 2D reference images RI1, RI2, RI3 and RI4 were respectively 15, 21.2, 7.5 and 7.8$\mu$m/pixel.

\begin{figure}[h]
	\centering
	\includegraphics[scale=.35]{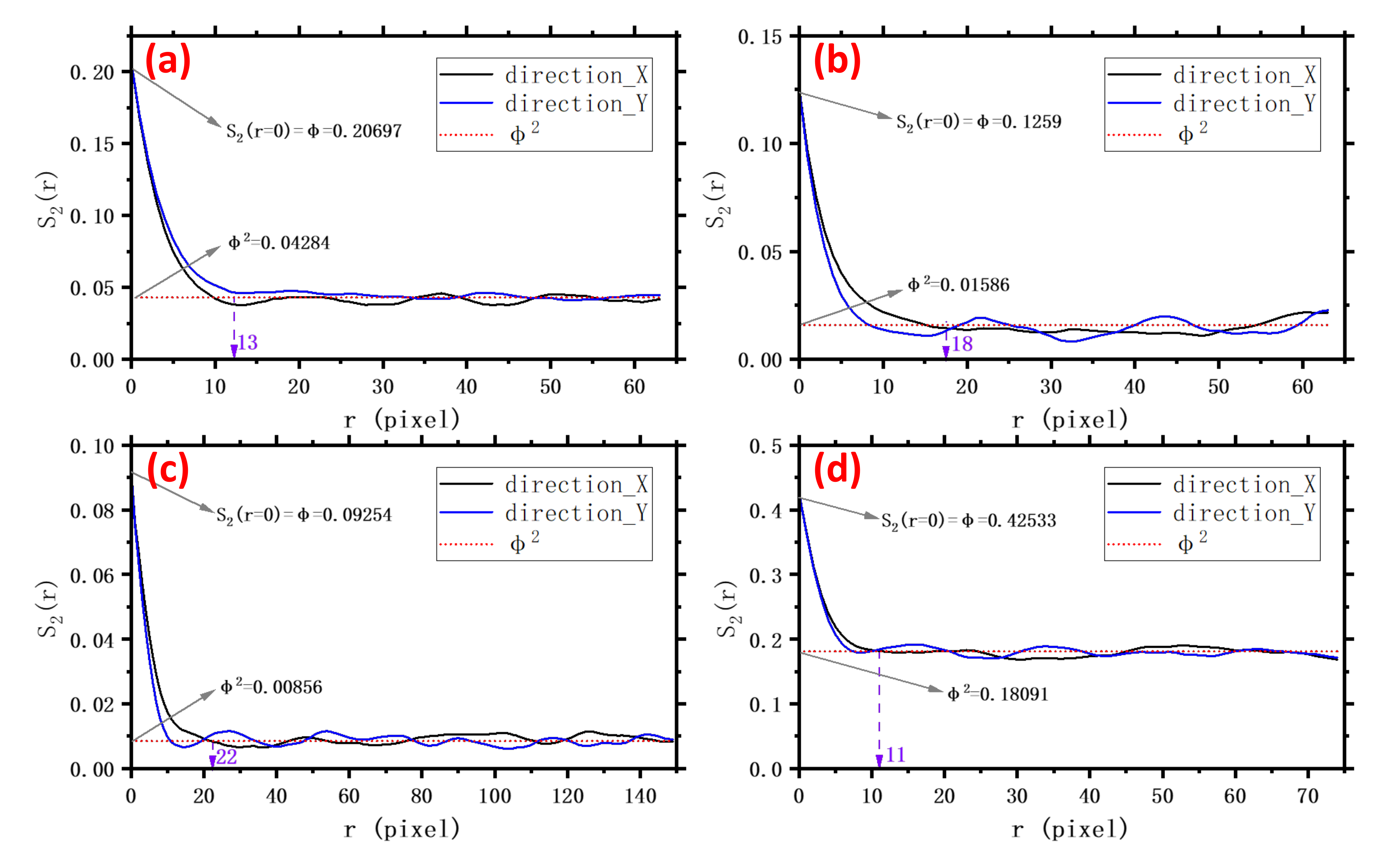}
	\caption{Two-point probability functions $S_2(r)$ and autocorrelation distances $l_{cor}$ of different reference images. (a)-(d) correspond to the reference images RI1-RI4, respectively.}
	\label{FIG:16}
\end{figure}

The structural parameters are shown in Figure \ref{FIG:16}. For Fontainebleau sandstone RI1, the porosity was 0.207, and the autocorrelation distance was estimated to be approximately 13. According to Equation \ref{Eq:7}, the depth of the network $m$ should be 6. Based on the previous experiment, the width of network $n$ was set to 16. Therefore, the network structure designed for RI1 is six layers with 16 channels ($L6C16$). For Ketton limestone RI2, the porosity was 0.126, and the autocorrelation distance was approximately 18. Similarly, the network structure used was nine layers with 16 channels ($L9C16$) using Equation \ref{Eq:7}. For silica material RI3, the porosity was 0.0925, and the autocorrelation distance was approximately 22. A network structure of 11 layers with 16 channels ($L11C16$) was used. The porosity was 0.425 for the synthetic ceramic material RI4, and the autocorrelation distance was approximately 11. Thus, the network structure was five layers with 16 channels ($L5C16$). 

\begin{figure}[h]
	\centering
	\includegraphics[scale=.12]{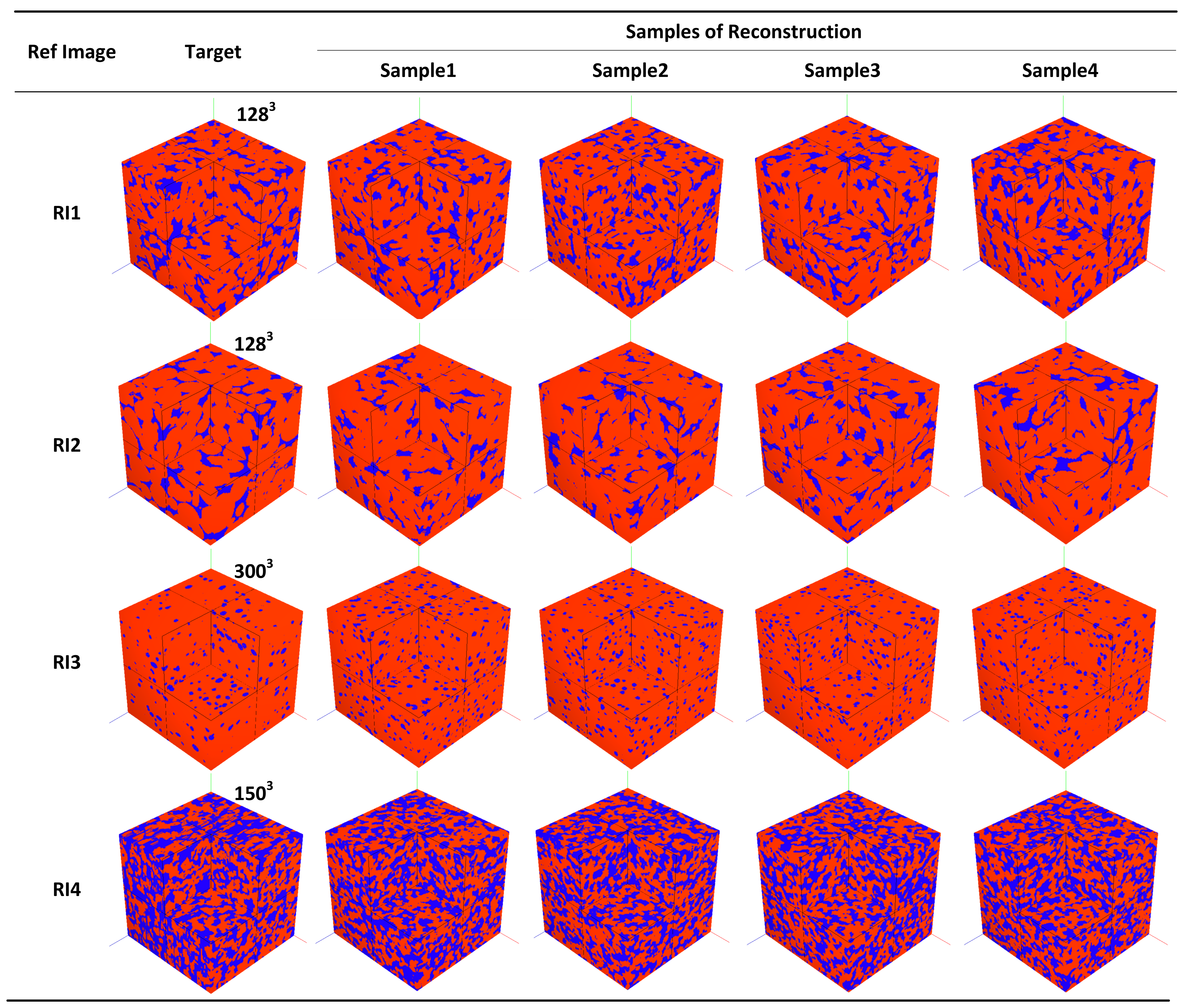}
	\caption{Comparison of 3D visualization between reconstrution results and corresponding target structure on RI1-RI4.}
	\label{FIG:17}
\end{figure}

For each reference image, four models were trained according to the corresponding network structure. The optimization process parameters for all models were set as a learning rate of 0.1, batch size of 1, and the number of iterations of 1000. In addition, four reconstructions were performed for each trained model, and 16 reconstruction results were obtained. A 3D structure was extracted from the four reconstruction results of each group for visualization; the 3D visualization results are shown in Figure \ref{FIG:17}. As in the previous experiment, a small piece was cut at the vertex of each reconstruction result to observe the internal structure. The first column is the real 3D structure corresponding to the reference image RI1–RI4, and the second to the fifth columns show the reconstruction results of the 16 different trained models. Moreover, the models corresponding to each row had the same network structure and training parameter settings, except for the initial random seeds of the network.

As shown in Figure \ref{FIG:17}, for the target 3D structure of the reference images RI1–RI4, there are some differences in the pore size and pore space morphology. However, when comparing the four reconstructed structures in each transverse row, the reconstructed structures reproduced almost the same pore size and pore space morphology as the target structures. This result shows that the $LmCn$ network, designed according to the structural information extracted from different reference images, establishes such a structural relationship. Regarding pore morphology, the selected description function of the VGG network can provide a relatively accurate description for different reference image morphologies. Through the visualization of the 3D structure, the reconstruction results successfully reproduced the morphological characteristics of the reference image, which proves the effectiveness of the proposed method in designing networks to a certain extent.

\begin{figure}[h]
	\centering
	\includegraphics[scale=.25]{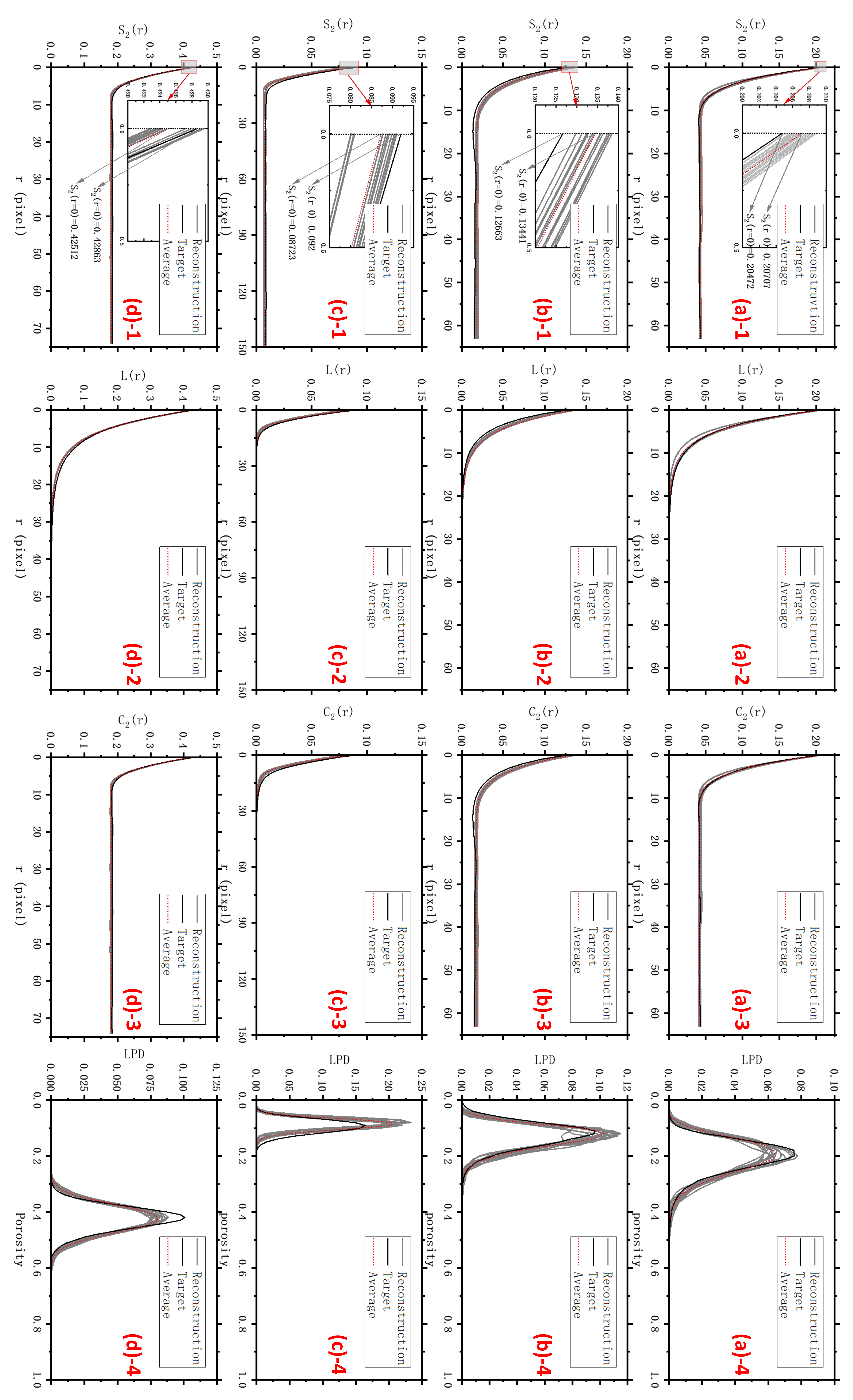}
	\caption{Comparison of statistical characteristic functions (two-point probability function $S_2(r)$, linear path function $L(r)$, two-point cluster function $C_2(r)$, and local porosity distribution $LPD$) between reconstruction results and the target structure on RI1-RI4. (a)-1 to (a)-4 are the statistical characteristic functions on RI1. (b)-1 to (b)-4 are the statistical characteristic functions on RI2. (c)-1 to (c)-4 are the statistical characteristic functions on RI3. (d)-1 to (d)-4 are the statistical characteristic functions on RI4.}
	\label{FIG:18}
\end{figure}

To accurately measure the similarity between the target 3D structure and the reconstruction results, the two-point correlation function \cite{9yeong1998reconstructing}, linear path function \cite{58lu1992lineal}, two-point cluster function \cite{59torquato1988two}, and local porosity distribution \cite{60biswal1998three} were used as evaluation indexes for statistical characteristics. The comparison results of the statistical characteristics are shown in Figure \ref{FIG:18}. For the two-point probability function, linear path function, and two-point cluster function, the target structures of the different reference images all showed large differences (Figure \ref{FIG:18}). In terms of the phase volume fraction, the porosity of the structure corresponding to RI4 was approximately 0.4, whereas that of the target structure corresponding to RI3 was less than 0.1. The same observation is true for the autocorrelation distance and linear path distribution. The maximum linear path length, or the maximum chord length, is close to 20 pixels for the target structures of RI1–RI3 but close to 30 pixels for RI4. Furthermore, the connectivity of the two points is quite different. The two points in the different connectivity domains of RI3 are almost separated, but the connectivity between the two points is good for the corresponding structures of RI1, RI3, and RI4. The spatial distribution of the pores also showed significant differences at the peak and broadening.

However, a lateral comparison of the reconstructed structures in each group showed that the basic statistical characteristics were relatively consistent. The curves of the two-point probability function, linear path function, and two-point cluster function of the reconstruction results were consistent with those of the target structure. It shows agreement in parameters such as porosity, autocorrelation distance, linear path distribution, and two-point connectivity, again proving that the designed network establishes the long-range structural relationship of the target structure. There was still a slight difference in the peak values for the local porosity distribution function. The peak values of RI1 and RI4 were slightly higher than those of the reconstructed structure, whereas the peak value of RI3 was slightly lower than that of the reconstructed structure. This result indicates that the reconstructed structure is still slightly different from the target structure in terms of overall homogeneity. However, the broadening is quite similar in the curves of the local porosity distribution function. This observation also illustrates that the reconstructed structure reproduced the spatial distribution of the pores of the target structure. In summary, the reconstruction of these four sets of reference images confirms the effectiveness and versatility of the proposed method.

\begin{figure}[h]
	\centering
	\includegraphics[scale=.35]{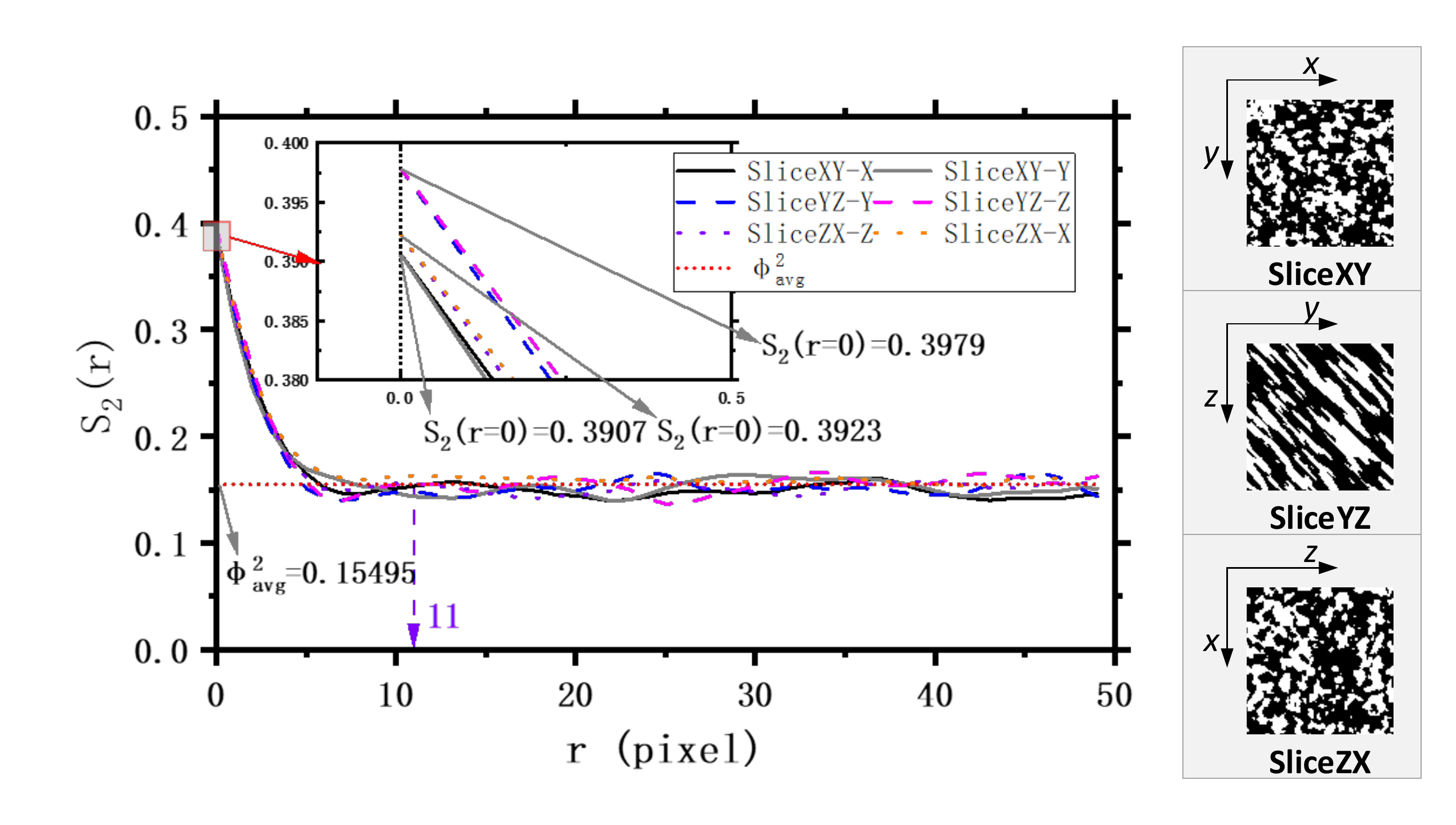}
	\caption{Reference images of the three slices used for the anisotropic material reconstruction. Two-point probability function $S_2(r)$ (left) and reference images (right).}
	\label{FIG:19}
\end{figure}

In addition, because the proposed method extracts three sections and characterizes the corresponding features with the description function as the optimization objective, the method should also have the corresponding reconstruction ability for some isotropic materials. To further verify the anisotropy reconstruction capability of the proposed method, an additional set of anisotropic materials \cite{61bostanabad2016characterization} was used in the validation experiments, as shown in Figure \ref{FIG:19}.

The anisotropy of the material shown in Figure \ref{FIG:19} is not shown in the X, Y, and Z directions but in other directions, and the porosities of the three images were 0.3907, 0.3979, and 0.3923, respectively, and the image size was 100. The autocorrelation distance was approximately 11. The network structure designed according to the guiding principle should have five layers and 16 channels ($L5C16$). The other parameter settings were the same as those in the previous experiment.

\begin{figure}[h]
	\centering
	\includegraphics[scale=.15]{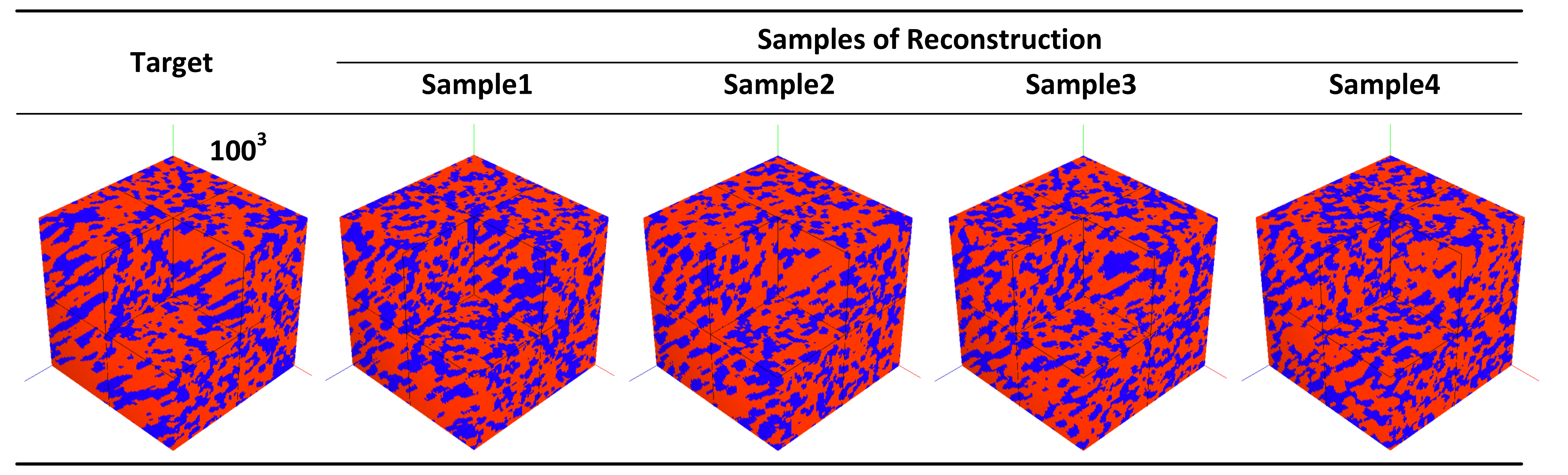}
	\caption{Comparison of 3D visualization between reconstruction results and the target structure on anisotropic slices.}
	\label{FIG:20}
\end{figure}

The experimental results were compared using subjective visual evaluations and objective statistical parameters. The results of the 3D visualization are shown in Figure \ref{FIG:20}. As shown in Figure \ref{FIG:20}, the reconstruction results show anisotropy in the same direction as the target structure. However, because its anisotropy is not shown in the X, Y, and Z directions, the structural parameters calculated using the three directions do not refer to the size of the structural parameters in the anisotropic direction. Therefore, it is visually shown that there is a slight difference in the chord length in the anisotropic direction between the target structure and reconstruction results.

\begin{figure}[h]
	\centering
	\includegraphics[scale=.35]{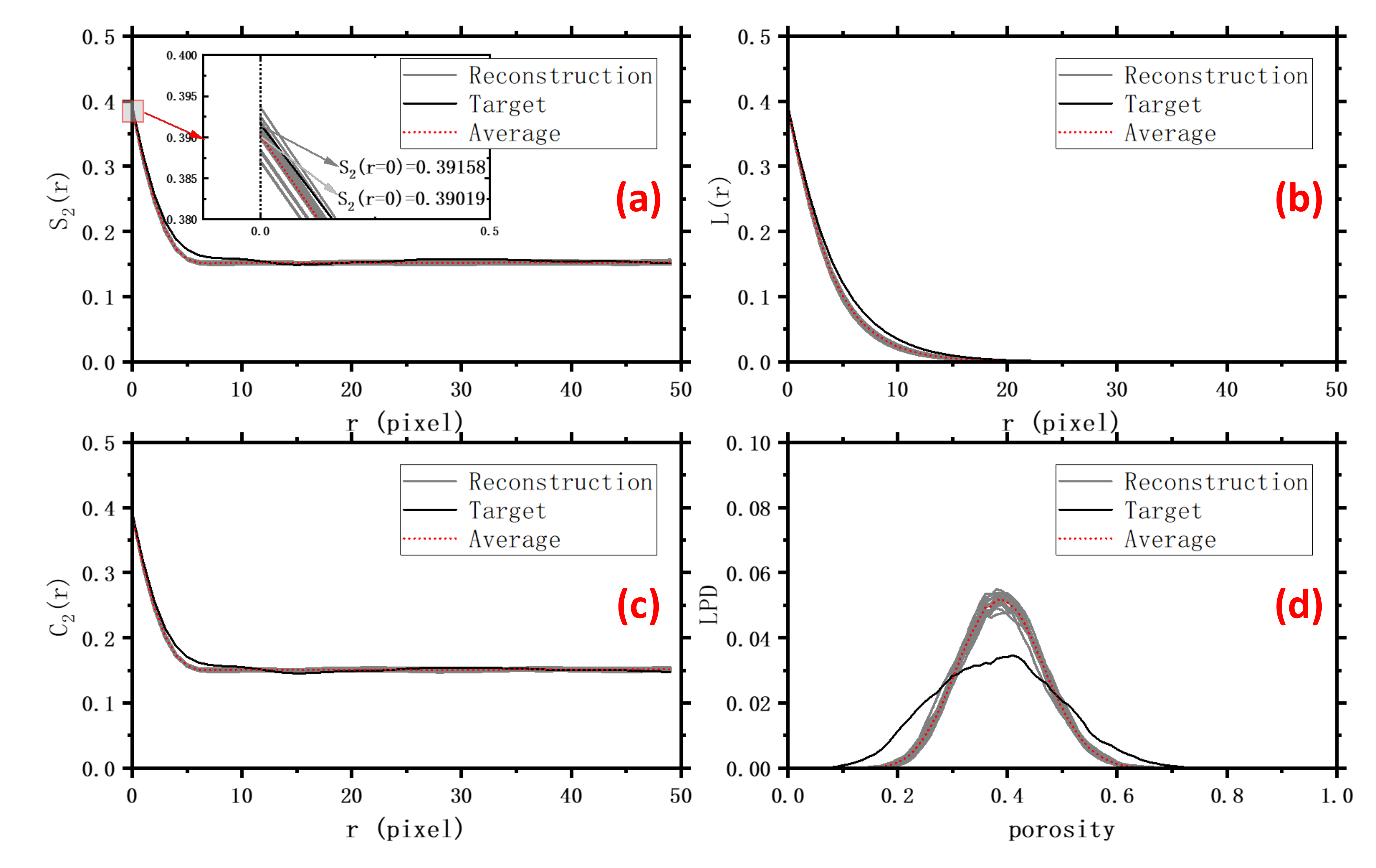}
	\caption{Comparison of statistical characteristic functions between reconstruction results and the target structure on anisotropic slices. (a) Two-point probability function $S_2(r)$. (b) Linear path function $L(r)$. (c) Two-point cluster function $C_2(r)$; (d) Local porosity distribution $LPD$.}
	\label{FIG:21}
\end{figure}

The results of the objective statistical parameters are shown in Figure \ref{FIG:21}. Because the two-point probability function, linear path function, and two-point cluster function only calculate the X, Y, and Z directions, and the structural parameters of the network also refer to the information of these three directions, the parameter curves of the three statistical functions of the reconstruction result are consistent with the curves of the target 3D structure. However, some differences were present in the local porosity distribution, both peak and broadening. The broadening of the target 3D structure is greater than that of the reconstructed structure, which indicates that the reconstructed structure lacks some large pore regions and some small pore regions of the target structure in the entire pore space distribution. However, because the two structures are essentially the same in porosity, the peak value of the local porosity distribution of the target structure is lower than that of the reconstructed structure. This finding also illustrates the previous visual observation results that the chord length of the reconstructed structure in the anisotropic direction is inferior to that of the target structure.

The results in both visualization and statistical parameters show that the proposed method can indeed carry out 3D reconstruction using a representative image in each of the three directions, and the reconstruction results can also show anisotropy. The proposed method has a specific reconstruction ability for a material whose three-directional sections can characterize the anisotropy. However, relative improvements are also necessary, such as calculating the structural parameters in more directions and choosing an appropriate one to design the network

\subsection{Verification of reconstruction in a large size}

For 3D reconstruction methods, reconstructing large reference images has always been difficult. As the size of the reference image increases, its feature distribution becomes more complex, and it becomes increasingly challenging to establish a map to achieve such a reconstruction. However, an increase in the reconstruction size will also lead to the problem of reduced reconstruction efficiency. Like traditional methods, it takes hours or even days to reconstruct the 512-size structure. However, deep learning methods also cause memory overhead. The number of intermediate result parameters of the running model is too large, which may lead to failure to train. They have also been studied in traditional and deep learning methods to adapt 3D reconstruction methods to increasingly complex reference images.

\begin{figure}[h]
	\centering
	\includegraphics[scale=.15]{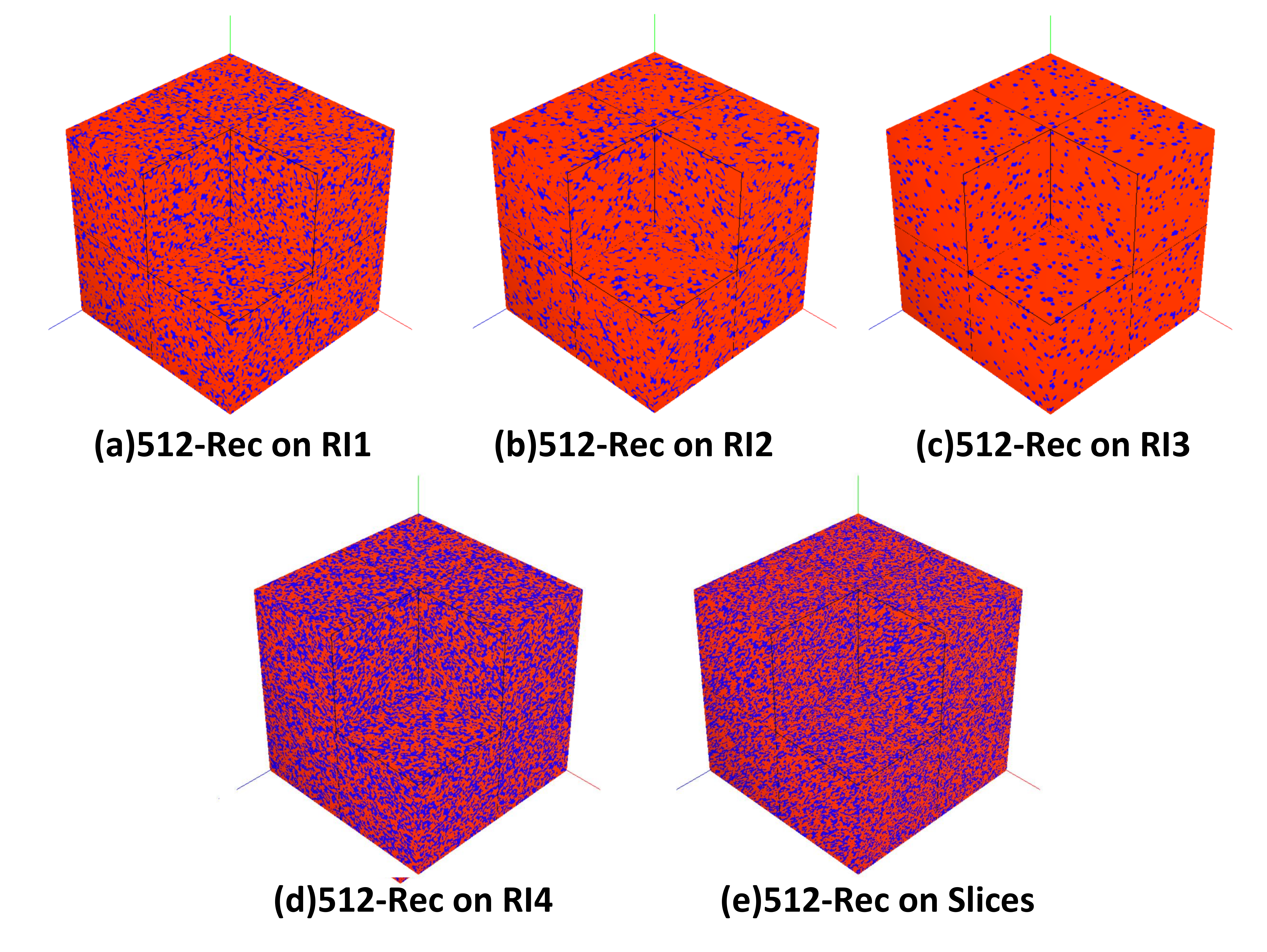}
	\caption{3D Visualization of large-size reconstructed structures. (a) 512-size reconstructed structure on RI1. (b) 512-size reconstructed structure on RI2. (c) 512-size reconstructed structure on RI3. (d) 512-size reconstructed structure on RI4. (f) 512-size reconstructed structure on anisotropic slices.}
	\label{FIG:22}
\end{figure}

In contrast to other methods, the model trained in this study can reconstruct images of any size with sufficient memory. In the case of limited memory, a large size can be reconstructed by cutting into sub-blocks. For each material image in Section 3.2, a 512-size structure was reconstructed separately by setting the sub-block size to 64, as shown in Figure \ref{FIG:22}. The potential advantage of the proposed method for large-size reference image reconstruction is that if the representative element area (REA) of the reference image is smaller than the size of the reference image itself. The REA-size image can be used for training in the optimization process, and only the noise structure input of the reference image size can be initialized during the reconstruction process.

Because the proposed method must establish a model for each reference image, the training time is also crucial to the reconstruction efficiency. Theoretically, the optimization time is expected to be as close as possible to reconstructing a 3D structure or even shorter than traditional methods because the proposed method adopts an improved framework similar to the SA method. 

All the experiments in Sections 3.2 and 3.3 were carried out on a computer configured with GTX1050Ti. In the optimization process, the  relevant parameters were set to be the same, learning rate was 0.1, batch size was 1, and number of iterations were 1000. In the reconstruction process, the sub-block sizes of RI1 and RI2 were 128, that of RI3 was 100, that of RI4 was 150, and that for the anisotropic material image was 100. All material image sub-blocks were 64 when reconstructing the corresponding 3D structure of size 512. The corresponding times of the optimization and reconstruction processes for four homogeneous materials and one anisotropic material image are shown in Table \ref{tbl1}.

\begin{table}
	\centering
	\caption{Time analysis of optimization and reconstruction process.}
	\label{tbl1}
	\begin{tabular}{llllll}
		\hline Time & RI1 & RI2 & RI3 & RI4 & Anisotropic slices \\
		\hline Average time of optimazition(s) & 507.18 & 983.31 & 4138.85 & 550.80 & 254.16 \\
		Average time of renconstrution(s) & 3.58 & 3.41 & 17.49 & 3.26 & 2.76 \\
		Time of rectructing in large size(s) & 40.48 & 67.18 & 88.49 & 33.66 & 33.05 \\
		\hline
	\end{tabular}
\end{table}

From Table \ref{tbl1}, for RI1, RI4, and anisotropic material images, because the structural parameters of the image itself are small, the network structure is simple, and only a few layers of the network are needed; thus, the training speed is fast, and it only takes about 10 min for optimization. For image RI2, which has large structural parameters and is relatively more complex, a 9-layer network is required, and the training time is also relatively long, but the optimization process can also be completed within 20 min. However, for RI3, the structural parameters are larger, and a deeper network is required to establish the mapping relationship. At the same time, the image size is much larger than the other images (size 300). Thus, the optimization process takes more than an hour to finish.

However, once the mapping is established, the time to complete the reconstruction process is less. Except for RI3, which has a relatively large reconstruction size and an average reconstruction time of 17.49 seconds, the rest of the reconstruction processes took approximately 3 s, where the reconstruction efficiency was much higher than that of  most traditional methods. Moreover, the proposed method can complete the reconstruction relatively quickly, even when a large-size reconstruction is performed. For a reconstructed structure of size 512, it takes less than 2 min to complete, which is also relatively efficient.

\section{Discussion}

The proposed method in this paper has some advantages in reconstruction efficiency, flexibility and generalization. Compared with the traditional methods, the proposed method obtains a mapping after optimization process, which can complete multiple groups of microstructure reconstruction in a short time. While compared with other deep learning methods, a key point of our method is that it cannot rely on 3D structure as training samples. In addition, our method can design different networks according to reference images. For some reference images with less complex structure, the designed network is simpler than networks of other deep learning methods, which speeds up its optimization process. As for images with more complex structure, the designed network can also be adjusted accordingly, which has more generalization than other networks theoretically. It is worth noting that the generalization here is not to train a universal model that can reconstruct the 3D structure of all kinds of images, but to stably train an effective network for different reference images. In addition, our method does not involve the adversative max-min game like GAN, but a directional optimization process to minimize the difference in the distribution of the features of the two images extracted by description function. Thus, the whole process will be more stable.

\begin{figure}[h]
	\centering
	\includegraphics[scale=.15]{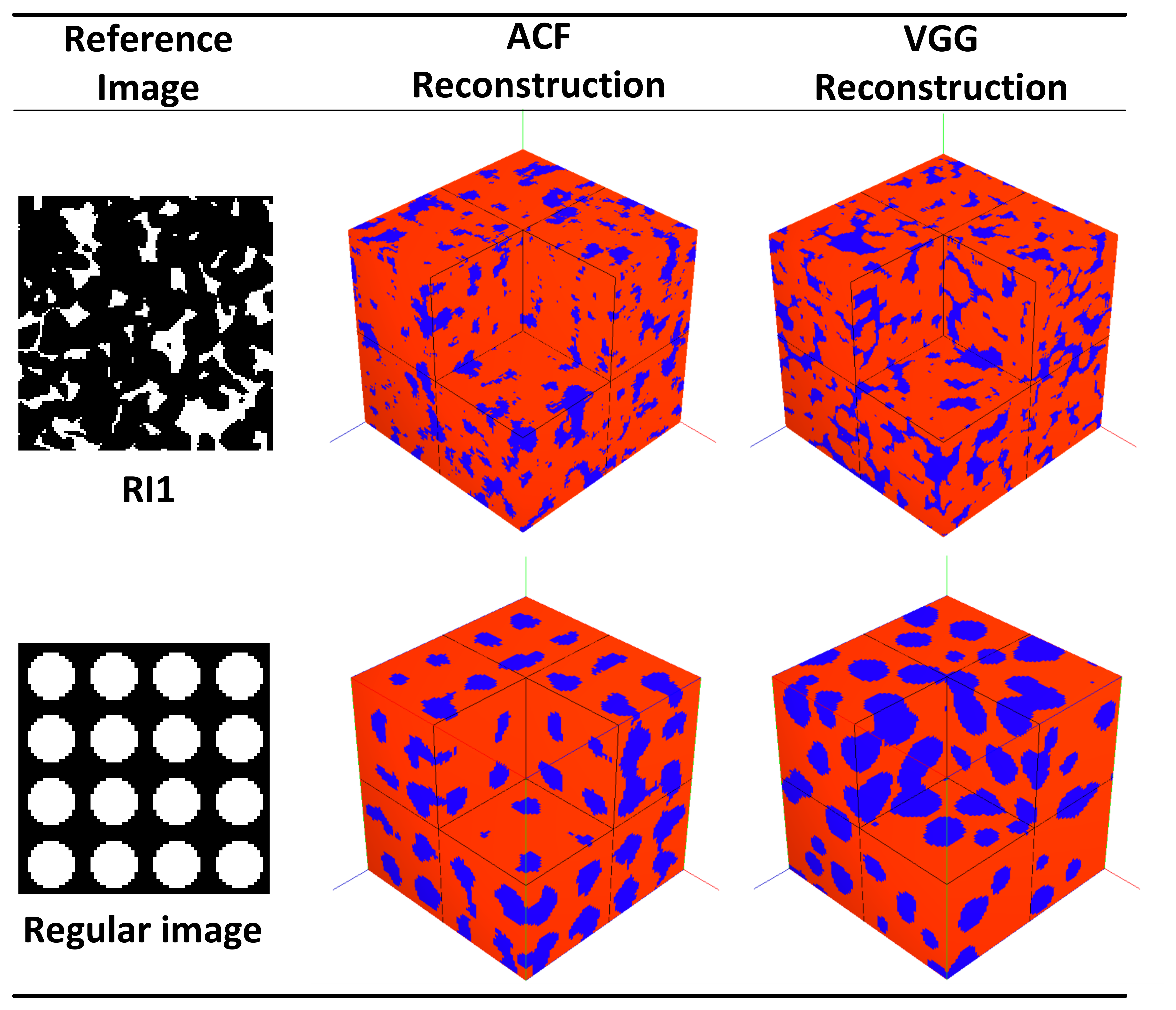}
	\caption{3D Visualization of reconstructed structures using different description functions.}
	\label{FIG:23}
\end{figure}

The description function is also an essential part of our method. In the proposed method, a statistical feature function, autocorrelation function (ACF), is designed for the description function. The calculation of ACF description function is given by Equation \ref{Eq:19}. Where $S_k$ is the autocorrelation function of the $Slice_k$, as shown in Equation \ref{Eq:20}; $R$ is the autocorrelation function of the reference image $Ref$, as shown in Equation \ref{Eq:21}. The reconstruction results under ACF can only satisfy the corresponding statistical parameters, and are less effective than those under VGG network in morphology. The results are shown in the Figure \ref{FIG:23}. However, ACF also has advantages in some aspects. In regular structures, it can play a better effect than the VGG network, as shown in Figure \ref{FIG:23}. The primary cause is that the VGG network characterizes more style and content features, and is slightly inferior to statistical functions in some specific structural features.

\begin{equation}\label{Eq:19}
	{\cal L}_{ACF}  = \sum\limits_k {\left\| {S_k  - R} \right\|} 
\end{equation}
\begin{equation}\label{Eq:20}
	R(\tau ,\upsilon ) = \sum\limits_{i =  - \infty }^{ + \infty } {\sum\limits_{j =  - \infty }^{ + \infty } {Ref(i,j)Ref(i + \tau ,j + \upsilon )} } 
\end{equation}
\begin{equation}\label{Eq:21}
	S_k (\tau ,\upsilon ) = \sum\limits_{i =  - \infty }^{ + \infty } {\sum\limits_{j =  - \infty }^{ + \infty } {Slice_k (i,j)Slice_k (i + \tau ,j + \upsilon )} } 
\end{equation}

However, the proposed method also has some limitations. Firstly, there is no quantitative description of the relationship between the width of the network and the complexity of the image. Although the parameter setting $n=16$ can be used in most cases, there should be a corresponding value for reference images of different complexity. The setting of this parameter is also relatively important, because it can further improve the network design. It can also reduce memory requirements and improve efficiency to a certain extent. 

Subsequently, VGG network and ACF are the only choices for description functions. Both of two description functions have certain limitations for the regular structure. In addition, it is difficult to reconstruct the heterogeneous microstructure only by these description functions. The reconstruction for some regular and heterogeneous microstructures is an issue to be addressed in the future work.

Finally, for some images with large structural parameters, such as 1000 or even larger, the designed networks are memory demanding and inefficient to complete the optimization process. The reconstruction for these large-size images is also a follow-up line.

\section{Conclusion}

The computational reconstruction methods aim to complete fast and effective reconstruction and achieve considerable generalization. Traditional methods such as SA and MPS have generality, but cannot be as fast and efficient as deep learning methods. However, deep learning methods rely on 3D structure and cannot be flexible as traditional methods. In this paper, the proposed method addressed these limitations to a certain extent. Based on the reference image to guide the network design, a certain generalization is achieved; based on neural networks and corresponding optimization and reconstruction strategies, fast and efficient reconstruction is achieved. Extensive experiments on multiple material images also verify the effectiveness and generalization of our method. 

It is worth noting that our method can complete the optimization and reconstruction with only one 2D reference image. The proposed method makes full use of the internal information of the reference image, such as morphology and pattern distribution, and provides a certain referential value for other deep learning methods. Subsequently, we complete the network design under the guidance of structural parameters, and explore the influence of network depth and width on the reconstruction results, which has certain reference significance for the interpretability of deep learning. In addition, our method has significance for the reconstruction of large-size microstructures, which is also of considerable research value for subsequent scholars to promote related work. Finally, our method can achieve fast and efficient reconstruction, which is of great significance for the numerical analysis of microstructure.

\section*{Code availability section}
Name of the code: A fast and flexible algorithm for microstructure reconstruction combining simulated annealing and deep learning.

Contact: qzteng@scu.edu.cn

Program language: Python

Software required: Windows (64 bit), anaconda3, cuda 10.1, torch 1.7.1, torchversion 0.8.2

The source codes are available for downloading at the link: https://github.com/scu-stu-mzc/A-fast-and-flexible-algorithm-for-microstructure-reconstruction.git

\section*{Appendix A}

\begin{figure}[h]
	\centering
	\includegraphics[scale=.5]{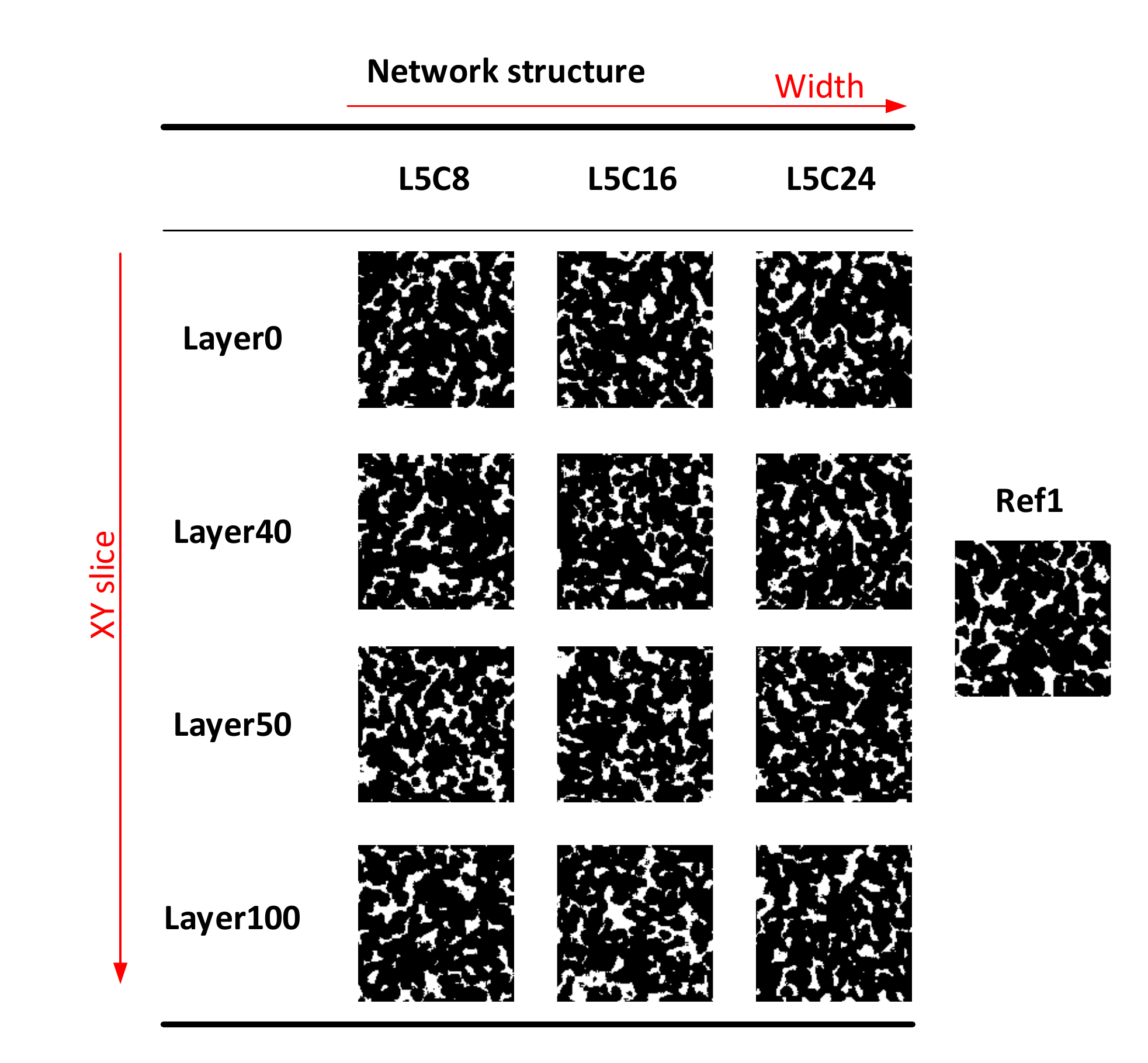}
	\caption{Comparison of 2D slices with different network widths on Ref1.}
	\label{FIG:A0}
\end{figure}

\begin{figure}[h]
	\centering
	\includegraphics[scale=.5]{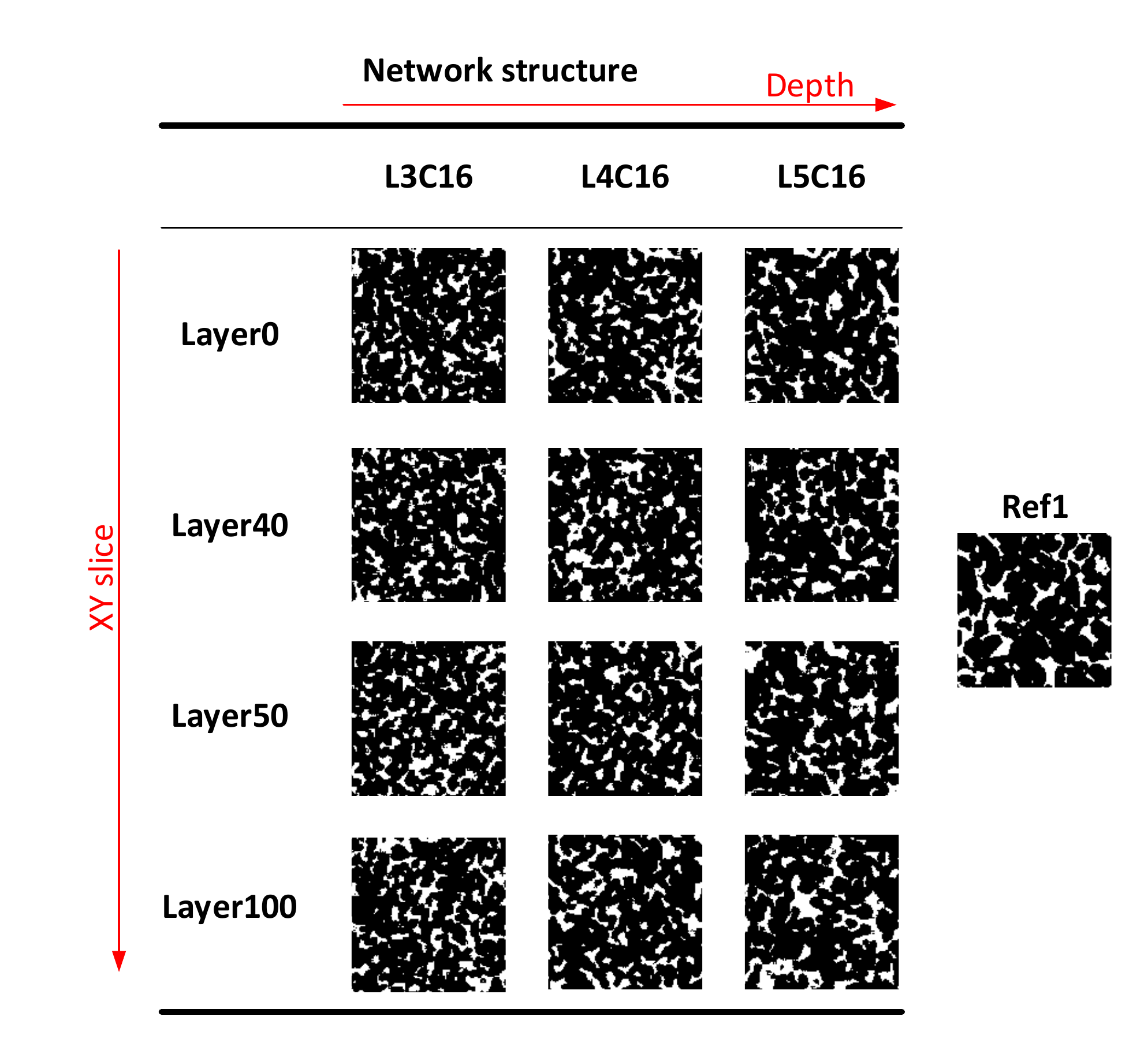}
	\caption{Comparison of 2D slices with different network depths on Ref1.}
	\label{FIG:A1}
\end{figure}

\begin{figure}[h]
	\centering
	\includegraphics[scale=.5]{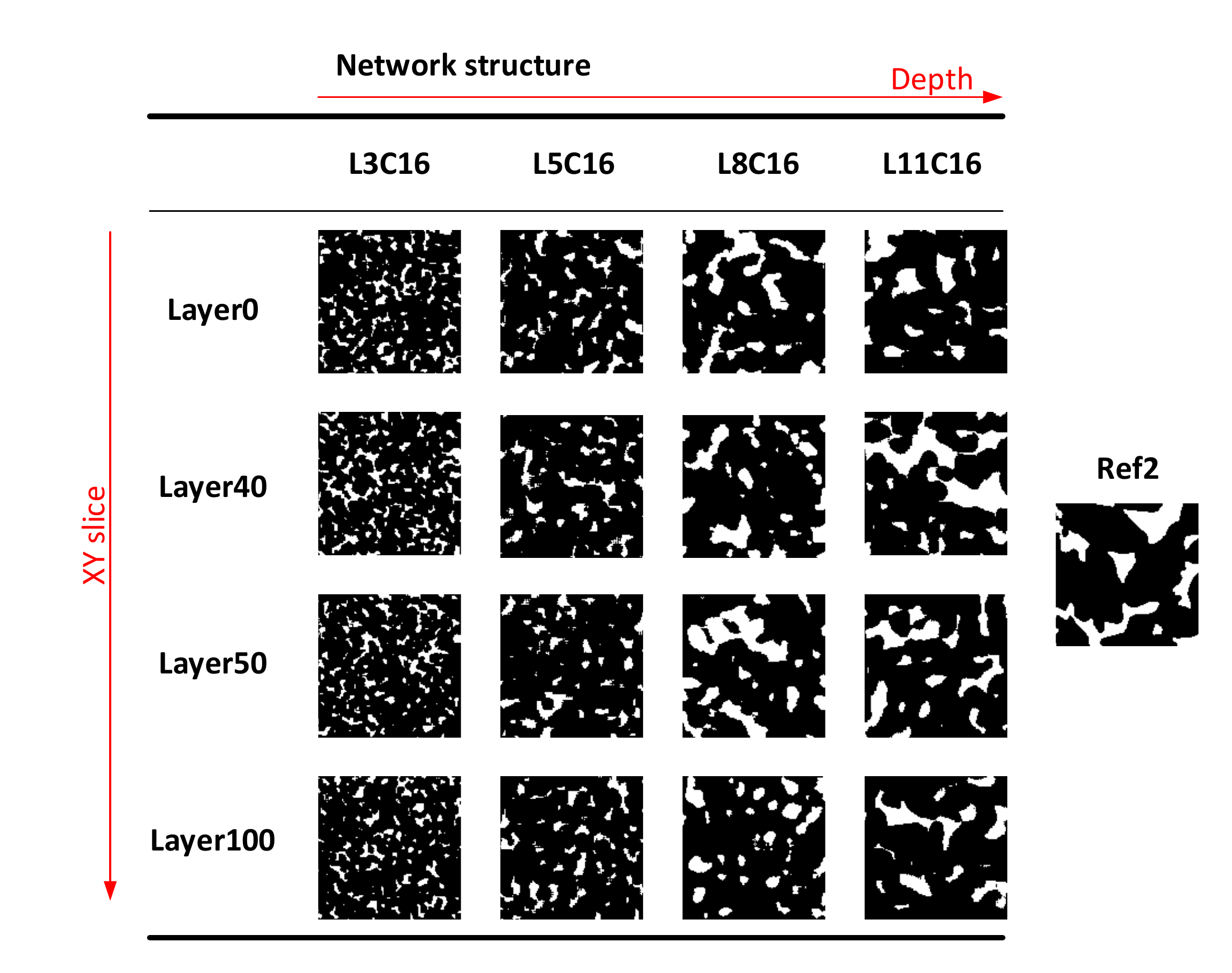}
	\caption{Comparison of 2D slices with different network depths on Ref2.}
	\label{FIG:A2}
\end{figure}

\begin{figure}[h]
	\centering
	\includegraphics[scale=.25]{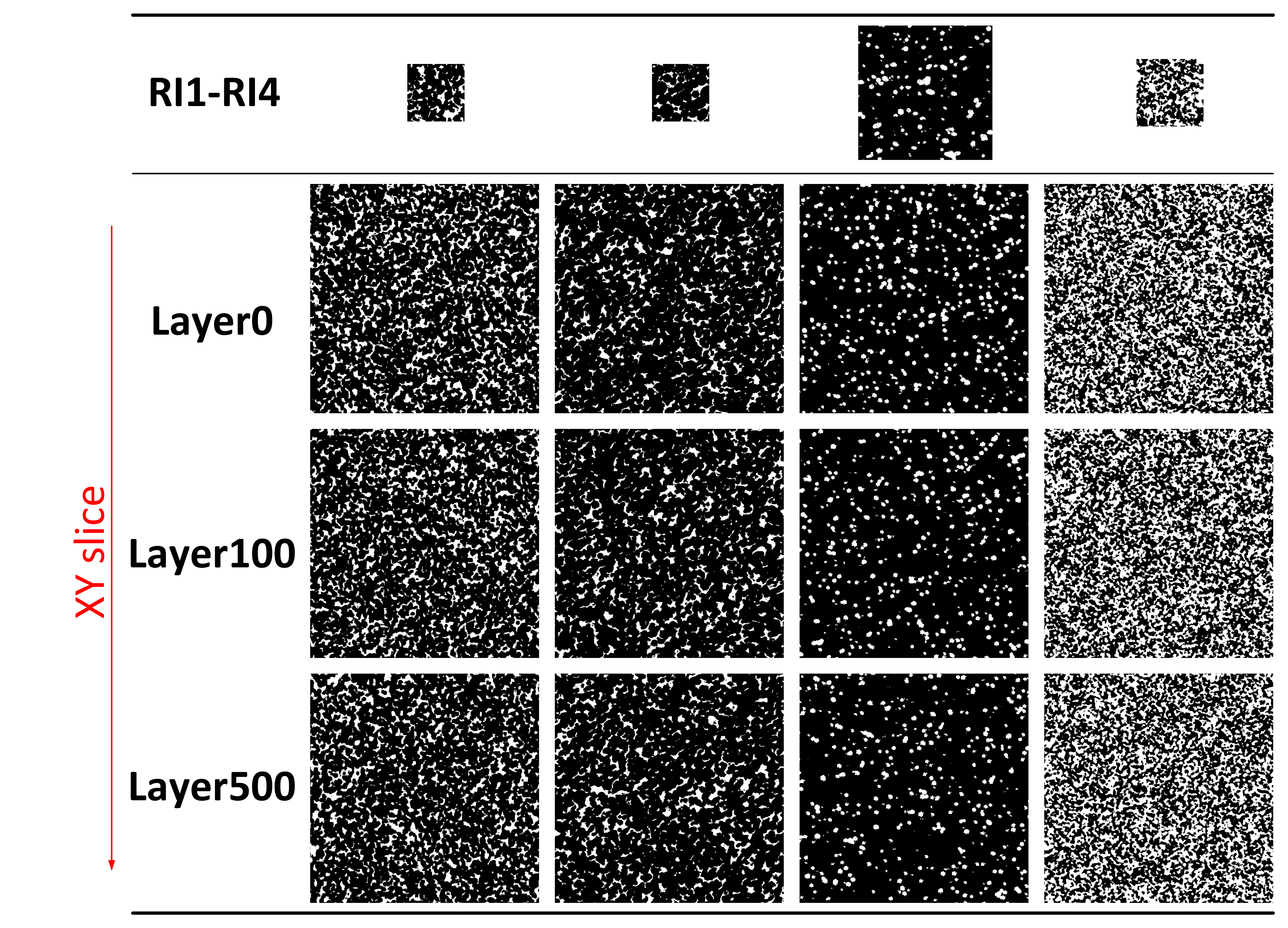}
	\caption{Comparison of 2D slices in large-size reconstruction on RI1-RI4.}
	\label{FIG:A3}
\end{figure}

\begin{figure}[h]
	\centering
	\includegraphics[scale=.25]{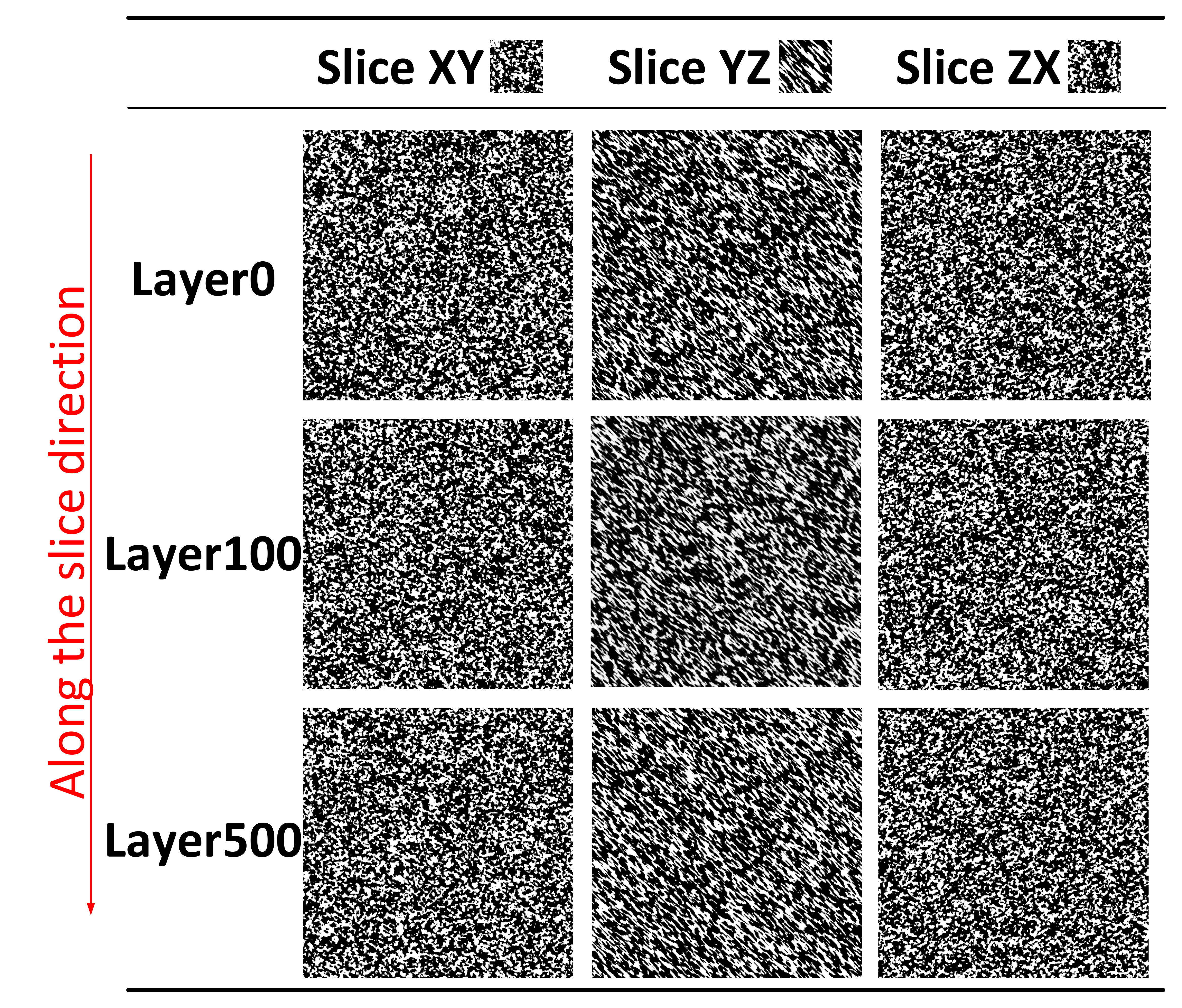}
	\caption{Comparison of 2D slices in large-size reconstruction on anisotropic slices.}
	\label{FIG:A4}
\end{figure}

In Appendix A, the experimental results in Section 3 are supplemented to make some more intuitive displays, mainly in the comparison of 2D slices for reconstructed structures, as shown in Figure \ref{FIG:A0}-\ref{FIG:A4}.

\bibliography{cas-refs}

\begin{thebibliography}{10}
\expandafter\ifx\csname url\endcsname\relax
  \def\url#1{\texttt{#1}}\fi
\expandafter\ifx\csname urlprefix\endcsname\relax\def\urlprefix{URL }\fi
\expandafter\ifx\csname href\endcsname\relax
  \def\href#1#2{#2} \def\path#1{#1}\fi

\bibitem{1zhao2012review}
C.~Zhao, Review on thermal transport in high porosity cellular metal foams with
  open cells, International Journal of Heat and Mass Transfer 55~(13-14) (2012)
  3618--3632.

\bibitem{2wang2013recent}
W.~Wang, Q.~Luo, B.~Li, X.~Wei, L.~Li, Z.~Yang, Recent progress in redox flow
  battery research and development, Advanced Functional Materials 23~(8) (2013)
  970--986.

\bibitem{3lan2019review}
Y.~Lan, Z.~Yang, P.~Wang, Y.~Yan, L.~Zhang, J.~Ran, A review of microscopic
  seepage mechanism for shale gas extracted by supercritical co2 flooding, Fuel
  238 (2019) 412--424.

\bibitem{4chen2022pore}
L.~Chen, A.~He, J.~Zhao, Q.~Kang, Z.-Y. Li, J.~Carmeliet, N.~Shikazono, W.-Q.
  Tao, Pore-scale modeling of complex transport phenomena in porous media,
  Progress in Energy and Combustion Science 88 (2022) 100968.

\bibitem{5iwai2010quantification}
H.~Iwai, N.~Shikazono, T.~Matsui, H.~Teshima, M.~Kishimoto, R.~Kishida,
  D.~Hayashi, K.~Matsuzaki, D.~Kanno, M.~Saito, et~al., Quantification of sofc
  anode microstructure based on dual beam fib-sem technique, Journal of Power
  Sources 195~(4) (2010) 955--961.

\bibitem{6wildenschild2013x}
D.~Wildenschild, A.~P. Sheppard, X-ray imaging and analysis techniques for
  quantifying pore-scale structure and processes in subsurface porous medium
  systems, Advances in Water resources 51 (2013) 217--246.

\bibitem{7tahmasebi2012reconstruction}
P.~Tahmasebi, M.~Sahimi, Reconstruction of three-dimensional porous media using
  a single thin section, Physical Review E 85~(6) (2012) 066709.

\bibitem{8bostanabad2018computational}
R.~Bostanabad, Y.~Zhang, X.~Li, T.~Kearney, L.~C. Brinson, D.~W. Apley, W.~K.
  Liu, W.~Chen, Computational microstructure characterization and
  reconstruction: Review of the state-of-the-art techniques, Progress in
  Materials Science 95 (2018) 1--41.

\bibitem{8_1sahimi2021reconstruction}
M.~Sahimi, P.~Tahmasebi, Reconstruction, optimization, and design of
  heterogeneous materials and media: Basic principles, computational
  algorithms, and applications, Physics Reports 939 (2021) 1--82.

\bibitem{9yeong1998reconstructing}
C.~Yeong, S.~Torquato, Reconstructing random media, Physical review E 57~(1)
  (1998) 495.

\bibitem{10strebelle2002conditional}
S.~Strebelle, Conditional simulation of complex geological structures using
  multiple-point statistics, Mathematical geology 34 (2002) 1--21.

\bibitem{11mariethoz2010direct}
G.~Mariethoz, P.~Renard, J.~Straubhaar, The direct sampling method to perform
  multiple-point geostatistical simulations, Water Resources Research 46~(11)
  (2010).

\bibitem{12mosser2017reconstruction}
L.~Mosser, O.~Dubrule, M.~J. Blunt, Reconstruction of three-dimensional porous
  media using generative adversarial neural networks, Physical Review E 96~(4)
  (2017) 043309.

\bibitem{13yang2018new}
M.~Yang, A.~Nagarajan, B.~Liang, S.~Soghrati, New algorithms for virtual
  reconstruction of heterogeneous microstructures, Computer Methods in Applied
  Mechanics and Engineering 338 (2018) 275--298.

\bibitem{14zhang2019efficient}
W.~Zhang, L.~Song, J.~Li, Efficient 3d reconstruction of random heterogeneous
  media via random process theory and stochastic reconstruction procedure,
  Computer Methods in Applied Mechanics and Engineering 354 (2019) 1--15.

\bibitem{15gao2021ultra}
Y.~Gao, Y.~Jiao, Y.~Liu, Ultra-efficient reconstruction of 3d microstructure
  and distribution of properties of random heterogeneous materials containing
  multiple phases, Acta Materialia 204 (2021) 116526.

\bibitem{16guo2023spherical}
F.-q. Guo, H.~Zhang, Z.-j. Yang, Y.-j. Huang, P.~J. Withers, A spherical
  harmonic-random field coupled method for efficient reconstruction of ct-image
  based 3d aggregates with controllable multiscale morphology, Computer Methods
  in Applied Mechanics and Engineering 406 (2023) 115901.

\bibitem{17gao2016pattern}
M.~Gao, Q.~Teng, X.~He, C.~Zuo, Z.~Li, Pattern density function for
  reconstruction of three-dimensional porous media from a single
  two-dimensional image, Physical Review E 93~(1) (2016) 012140.

\bibitem{18karsanina2018hierarchical}
M.~V. Karsanina, K.~M. Gerke, Hierarchical optimization: Fast and robust
  multiscale stochastic reconstructions with rescaled correlation functions,
  Physical review letters 121~(26) (2018) 265501.

\bibitem{19feng2018reconstruction}
J.~Feng, Q.~Teng, X.~He, L.~Qing, Y.~Li, Reconstruction of three-dimensional
  heterogeneous media from a single two-dimensional section via co-occurrence
  correlation function, Computational Materials Science 144 (2018) 181--192.

\bibitem{20zhou20183d}
X.-P. Zhou, N.~Xiao, 3d numerical reconstruction of porous sandstone using
  improved simulated annealing algorithms, Rock Mechanics and Rock Engineering
  51 (2018) 2135--2151.

\bibitem{21zhang2019high}
Y.~Zhang, M.~Yan, Y.~Wan, Z.~Jiao, Y.~Chen, F.~Chen, C.~Xia, M.~Ni,
  High-throughput 3d reconstruction of stochastic heterogeneous microstructures
  in energy storage materials, npj Computational Materials 5~(1) (2019) 11.

\bibitem{22alexander2009hierarchical}
S.~K. Alexander, P.~Fieguth, M.~A. Ioannidis, E.~R. Vrscay, Hierarchical
  annealing for synthesis of binary images, Mathematical geosciences 41 (2009)
  357--378.

\bibitem{23tang2009pixel}
T.~TANG, Q.-z. TENG, X.-h. HE, D.~Luo, A pixel selection rule based on the
  number of different-phase neighbours for the simulated annealing
  reconstruction of sandstone microstructure, Journal of microscopy 234~(3)
  (2009) 262--268.

\bibitem{24song2019improved}
S.~Song, An improved simulated annealing algorithm for reconstructing 3d
  large-scale porous media, Journal of Petroleum Science and Engineering 182
  (2019) 106343.

\bibitem{25xiao2022novel}
N.~Xiao, X.~Zhou, T.~Ling, Novel cooling--solidification annealing
  reconstruction of rock models, Acta Geotechnica (2022) 1--18.

\bibitem{26yang2018new}
M.~Yang, A.~Nagarajan, B.~Liang, S.~Soghrati, New algorithms for virtual
  reconstruction of heterogeneous microstructures, Computer Methods in Applied
  Mechanics and Engineering 338 (2018) 275--298.

\bibitem{27seibert2022descriptor}
P.~Seibert, A.~Ra{\ss}loff, M.~Ambati, M.~K{\"a}stner, Descriptor-based
  reconstruction of three-dimensional microstructures through gradient-based
  optimization, Acta Materialia 227 (2022) 117667.

\bibitem{28bagherian2022new}
A.~Bagherian, S.~Famouri, M.~Baghani, D.~George, A.~Sheidaei, M.~Baniassadi, A
  new statistical descriptor for the physical characterization and 3d
  reconstruction of heterogeneous materials, Transport in Porous Media
  142~(1-2) (2022) 23--40.

\bibitem{29chen2022fast}
D.~Chen, Z.~Xu, X.~Wang, H.~He, Z.~Du, J.~Nan, Fast reconstruction of
  multiphase microstructures based on statistical descriptors, Physical Review
  E 105~(5) (2022) 055301.

\bibitem{30ajani2022microstructural}
C.~K. Ajani, Z.~Zhu, D.-W. Sun, Microstructural classification and
  reconstruction of the computational geometry of steamed bread using
  descriptor-based approach, Transport in Porous Media 144~(2) (2022) 317--336.

\bibitem{31tahmasebi2014ms}
P.~Tahmasebi, M.~Sahimi, J.~Caers, Ms-ccsim: accelerating pattern-based
  geostatistical simulation of categorical variables using a multi-scale search
  in fourier space, Computers \& Geosciences 67 (2014) 75--88.

\bibitem{32pourfard2017pcto}
M.~Pourfard, M.~J. Abdollahifard, K.~Faez, S.~A. Motamedi, T.~Hosseinian,
  Pcto-sim: Multiple-point geostatistical modeling using parallel conditional
  texture optimization, Computers \& Geosciences 102 (2017) 116--138.

\bibitem{33zuo2020tree}
C.~Zuo, Z.~Yin, Z.~Pan, E.~J. MacKie, J.~Caers, A tree-based direct sampling
  method for stochastic surface and subsurface hydrological modeling, Water
  Resources Research 56~(2) (2020) e2019WR026130.

\bibitem{34gao2015reconstruction}
M.~Gao, X.~He, Q.~Teng, C.~Zuo, D.~Chen, Reconstruction of three-dimensional
  porous media from a single two-dimensional image using three-step sampling,
  Physical Review E 91~(1) (2015) 013308.

\bibitem{35gravey2020quicksampling}
M.~Gravey, G.~Mariethoz, Quicksampling v1. 0: a robust and simplified
  pixel-based multiple-point simulation approach, Geoscientific Model
  Development 13~(6) (2020) 2611--2630.

\bibitem{36bai2020hybrid}
T.~Bai, P.~Tahmasebi, Hybrid geological modeling: Combining machine learning
  and multiple-point statistics, Computers \& geosciences 142 (2020) 104519.

\bibitem{37wang2022two}
X.~Wang, S.~Yu, S.~Li, N.~Zhang, Two parameter optimization methods of
  multi-point geostatistics, Journal of Petroleum Science and Engineering 208
  (2022) 109724.

\bibitem{38li2018markov}
Y.~Li, X.~He, Q.~Teng, J.~Feng, X.~Wu, Markov prior-based block-matching
  algorithm for superdimension reconstruction of porous media, Physical Review
  E 97~(4) (2018) 043306.

\bibitem{39xia2021three}
Z.~Xia, Q.~Teng, X.~Wu, J.~Li, P.~Yan, Three-dimensional reconstruction of
  porous media using super-dimension-based adjacent block-matching algorithm,
  Physical Review E 104~(4) (2021) 045308.

\bibitem{40fu2022stochastic}
J.~Fu, D.~Xiao, D.~Li, H.~R. Thomas, C.~Li, Stochastic reconstruction of 3d
  microstructures from 2d cross-sectional images using machine learning-based
  characterization, Computer Methods in Applied Mechanics and Engineering 390
  (2022) 114532.

\bibitem{41feng2018accelerating}
J.~Feng, Q.~Teng, X.~He, X.~Wu, Accelerating multi-point statistics
  reconstruction method for porous media via deep learning, Acta Materialia 159
  (2018) 296--308.

\bibitem{42feng2020end}
J.~Feng, Q.~Teng, B.~Li, X.~He, H.~Chen, Y.~Li, An end-to-end three-dimensional
  reconstruction framework of porous media from a single two-dimensional image
  based on deep learning, Computer Methods in Applied Mechanics and Engineering
  368 (2020) 113043.

\bibitem{43valsecchi2020stochastic}
A.~Valsecchi, S.~Damas, C.~Tubilleja, J.~Arechalde, Stochastic reconstruction
  of 3d porous media from 2d images using generative adversarial networks,
  Neurocomputing 399 (2020) 227--236.

\bibitem{44zhang20213d}
T.~Zhang, P.~Xia, F.~Lu, 3d reconstruction of digital cores based on a model
  using generative adversarial networks and variational auto-encoders, Journal
  of Petroleum Science and Engineering 207 (2021) 109151.

\bibitem{45volkhonskiy2022generative}
D.~Volkhonskiy, E.~Muravleva, O.~Sudakov, D.~Orlov, E.~Burnaev, D.~Koroteev,
  B.~Belozerov, V.~Krutko, Generative adversarial networks for reconstruction
  of three-dimensional porous media from two-dimensional slices, Physical
  Review E 105~(2) (2022) 025304.

\bibitem{46zhang20223d}
T.~Zhang, X.~Ji, F.~Lu, 3d reconstruction of porous media by combining scaling
  transformation and multi-scale discrimination using generative adversarial
  networks, Journal of Petroleum Science and Engineering 209 (2022) 109815.

\bibitem{47henkes2022three}
A.~Henkes, H.~Wessels, Three-dimensional microstructure generation using
  generative adversarial neural networks in the context of continuum
  micromechanics, Computer Methods in Applied Mechanics and Engineering 400
  (2022) 115497.

\bibitem{48bostanabad2020reconstruction}
R.~Bostanabad, Reconstruction of 3d microstructures from 2d images via transfer
  learning, Computer-Aided Design 128 (2020) 102906.

\bibitem{49fu2021statistical}
J.~Fu, S.~Cui, S.~Cen, C.~Li, Statistical characterization and reconstruction
  of heterogeneous microstructures using deep neural network, Computer Methods
  in Applied Mechanics and Engineering 373 (2021) 113516.

\bibitem{50zhang20223d}
F.~Zhang, X.~He, Q.~Teng, X.~Wu, X.~Dong, 3d-pmrnn: Reconstructing
  three-dimensional porous media from the two-dimensional image with recurrent
  neural network, Journal of Petroleum Science and Engineering 208 (2022)
  109652.

\bibitem{51anderson2020rockflow}
T.~I. Anderson, K.~M. Guan, B.~Vega, S.~A. Aryana, A.~R. Kovscek, Rockflow:
  Fast generation of synthetic source rock images using generative flow models,
  Energies 13~(24) (2020) 6571.

\bibitem{52vgg}
K.~Simonyan, A.~Zisserman, Very deep convolutional networks for large-scale
  image recognition, in: International Conference on Learning Representations
  (ICLR), 2015, pp. 1--14.

\bibitem{532014Adam}
D.~Kingma, J.~Ba, Adam: A method for stochastic optimization, Computer Science
  (2014).

\bibitem{54gerke2018finite}
K.~M. Gerke, R.~V. Vasilyev, S.~Khirevich, D.~Collins, M.~V. Karsanina, T.~O.
  Sizonenko, D.~V. Korost, S.~Lamontagne, D.~Mallants, Finite-difference method
  stokes solver (fdmss) for 3d pore geometries: Software development,
  validation and case studies, Computers \& geosciences 114 (2018) 41--58.

\bibitem{55coker1996morphology}
D.~A. Coker, S.~Torquato, J.~H. Dunsmuir, Morphology and physical properties of
  fontainebleau sandstone via a tomographic analysis, Journal of Geophysical
  Research: Solid Earth 101~(B8) (1996) 17497--17506.

\bibitem{56dong2009pore}
H.~Dong, M.~J. Blunt, Pore-network extraction from
  micro-computerized-tomography images, Physical review E 80~(3) (2009) 036307.

\bibitem{57muljadi2016impact}
B.~P. Muljadi, M.~J. Blunt, A.~Q. Raeini, B.~Bijeljic, The impact of porous
  media heterogeneity on non-darcy flow behaviour from pore-scale simulation,
  Advances in water resources 95 (2016) 329--340.

\bibitem{58lu1992lineal}
B.~Lu, S.~Torquato, Lineal-path function for random heterogeneous materials,
  Physical Review A 45~(2) (1992) 922.

\bibitem{59torquato1988two}
S.~Torquato, J.~Beasley, Y.~Chiew, Two-point cluster function for continuum
  percolation, The Journal of chemical physics 88~(10) (1988) 6540--6547.

\bibitem{60biswal1998three}
B.~Biswal, C.~Manwart, R.~Hilfer, Three-dimensional local porosity analysis of
  porous media, Physica A: Statistical Mechanics and its Applications 255~(3-4)
  (1998) 221--241.

\bibitem{61bostanabad2016characterization}
R.~Bostanabad, W.~Chen, D.~Apley, Characterization and reconstruction of 3d
  stochastic microstructures via supervised learning, Journal of microscopy
  264~(3) (2016) 282--297.

\end{thebibliography}

\end{document}